\documentclass[twocolumn]{aastex63}
\usepackage{hyperref}
\usepackage[colorinlistoftodos]{todonotes}
\usepackage{amsmath}
\shorttitle{Stellar Activity Mitigation}

\usepackage[T1]{fontenc} 
\usepackage[utf8]{inputenc} 

\begin{document}
\newcommand{\vdag}{(v)^\dagger}
\newcommand\aastex{AAS\TeX}
\newcommand\latex{La\TeX}
\newcommand\cms{$\mathrm{cm\ s^{-1}}$}
\newcommand\ms{$\mathrm{m\ s^{-1}}$}
\newcommand\kms{$\mathrm{km\ s^{-1}}$}
\newcommand\HDtwo{HD~26965}
\newcommand{\PSUAA}{Department of Astronomy \& Astrophysics, 525 Davey Laboratory, The Pennsylvania State University, University Park, PA, 16802, USA}
\newcommand{\PSUCEHW}{Center for Exoplanets and Habitable Worlds, 525 Davey Laboratory, The Pennsylvania State University, University Park, PA, 16802, USA}
\newcommand{\PSETI}{Penn State Extraterrestrial Intelligence Center, 525 Davey Laboratory, The Pennsylvania State University, University Park, PA, 16802, USA}
\newcommand{\UA}{Steward Observatory, The University of Arizona, 933 N.\ Cherry Ave, Tucson, AZ 85721, USA}
\newcommand{\UAA}{Department of Astronomy and Steward Observatory, University of Arizona, Tucson, AZ 85721, USA}
\newcommand{\Penn}{Department of Physics and Astronomy, University of Pennsylvania, 209 S 33rd St, Philadelphia, PA 19104, USA}
\newcommand{\Caltech}{Department of Astronomy, California Institute of Technology, Pasadena, CA 91125, USA}
\newcommand{\STScI}{Space Telescope Science Institute, 3700 San Martin Dr, Baltimore, MD 21218, USA}
\newcommand{\JHU}{Department of Physics and Astronomy, Johns Hopkins University, 3400 N Charles St, Baltimore, MD 21218, USA}
\newcommand{\GoddardESAL}{Exoplanets and Stellar Astrophysics Laboratory, NASA Goddard Space Flight Center, Greenbelt, MD 20771, USA}
\newcommand{\GoddardISTD}{Instrument Systems and Technology Division, NASA Goddard Space Flight Center, Greenbelt, MD 20771, USA}
\newcommand{\GSFC}{NASA Goddard Space Flight Center, Greenbelt, MD 20771, USA}
\newcommand{\NOAO}{NSF's National Optical-Infrared Astronomy Research Laboratory, 950 N.\ Cherry Ave., Tucson, AZ 85719, USA}
\newcommand{\UW}{Wisconsin address goes here}
\newcommand{\Macquarie}{School of Mathematical and Physical Sciences, Macquarie University, Balaclava Road, North Ryde, NSW 2109, Australia }
\newcommand{\NIST}{National Institute of Standards \& Technology, 325 Broadway, Boulder, CO 80305, USA}
\newcommand{\CUBoulder}{Department of Physics, 390 UCB, University of Colorado, Boulder, CO 80309, USA}
\newcommand{\JPL}{Jet Propulsion Laboratory, California Institute of Technology, 4800 Oak Grove Drive, Pasadena, California 91109}
\newcommand{\MIT}{Kavli Institute for Astrophysics and Space Research, Massachusetts Institute of Technology, Cambridge, MA, USA}
\newcommand{\UCI}{Department of Physics \& Astronomy, The University of California, Irvine, Irvine, CA 92697, USA}
\newcommand{\Carleton}{Carleton College, One North College St., Northfield, MN 55057, USA}
\newcommand{\Carnegie}{Earth and Planets Laboratory, Carnegie Institution for Science, 5241 Broad Branch Road, NW, Washington, DC 20015, USA}
\newcommand{\PSUICS}{Institute for Computational and Data Sciences, The Pennsylvania State University, University Park, PA, 16802, USA}
\newcommand{\PSUCASt}{Center for Astrostatistics, 525 Davey Laboratory, The Pennsylvania State University, University Park, PA, 16802, USA}
\newcommand{\NESSF}{NASA Earth and Space Science Fellow}
\newcommand{\Princeton}{Department of Astrophysical Sciences, Princeton University, 4 Ivy Lane, Princeton, NJ 08540, USA}
\newcommand{\RUSSELL}{Henry Norris Russell Fellow}
\newcommand{\IAS}{Institute for Advance Study, 1 Einstein Drive, Princeton, NJ 08540, USA}
\newcommand{\Tsinghua}{Department of Astronomy, Tsinghua University, Beijing 100084, China}
\newcommand{\FlatironCCA}{Center for Computational Astrophysics, Flatiron Institute, 162 Fifth Avenue, New York, NY 10010, USA}
\newcommand{\ETH}{ETH Zurich, Institute for Particle Physics \& Astrophysics, Zurich, Switzerland}
\newcommand{\Geneve}{Observatoire Astronomique de l’Université de Genève, Chemin Pegasi 51, 1290 Versoix, Switzerland}

\title{Quiet Please: Detrending Radial Velocity Variations from Stellar Activity with a Physically Motivated Spot Model}

\correspondingauthor{Jared Siegel}
\email{siegeljc@princeton.edu}

\author[0000-0002-9337-0902]{Jared C. Siegel}
\altaffiliation{NSF Graduate Research Fellow}
\affiliation{\Princeton}

\author[0000-0003-1312-9391]{Samuel Halverson}
\affil{\JPL}

\author[0000-0002-4927-9925]{Jacob K. Luhn}
\affil{\UCI}

\author[0000-0002-3852-3590]{Lily L. Zhao}
\affil{\FlatironCCA}

\author[0000-0002-3212-5778]{Khaled Al Moulla}
\affil{\Geneve}

\author[0000-0003-0149-9678]{Paul Robertson}
\affil{\UCI}

\author[0000-0003-4384-7220]{Chad F.\ Bender}
\affil{\UA}

\author[0000-0002-4788-8858]{Ryan C. Terrien}
\affil{\Carleton}

\author[0000-0001-8127-5775]{Arpita Roy}
\affil{\STScI}
\affil{\JHU}

\author[0000-0001-9596-7983]{Suvrath Mahadevan}
\affil{\PSUAA}
\affil{\PSUCEHW}

\author[0000-0002-1664-3102]{Fred Hearty}
\affil{\PSUAA}
\affil{\PSUCEHW}

\author[0000-0001-8720-5612]{Joe P.\ Ninan}
\affil{\PSUAA}
\affil{\PSUCEHW}

\author[0000-0001-6160-5888]{Jason T.\ Wright}
\affil{\PSUAA}
\affil{\PSUCEHW}
\affil{\PSETI}

\author[0000-0001-6545-639X]{Eric B.\ Ford}
\affil{\PSUAA}
\affil{\PSUCEHW}
\affil{\PSUICS}
\affil{\PSUCASt}

\author[0000-0002-4046-987X]{Christian Schwab}
\affil{\Macquarie}

\author[0000-0001-7409-5688]{Guðmundur Stefánsson} 
\altaffiliation{NASA Sagan Fellow}
\affil{Anton Pannekoek Institute for Astronomy, University of Amsterdam, Science Park 904, 1098 XH Amsterdam, The Netherlands}
\affil{Department of Astrophysical Sciences, Princeton University, 4 Ivy Lane, Princeton, NJ 08540, USA}


\author[0000-0002-6096-1749]{Cullen H.\ Blake}
\affil{\Penn}

\author[0000-0003-0241-8956]{Michael W.\ McElwain}
\affil{\GoddardESAL} 

\begin{abstract}
For solar-type stars, spots and their associated magnetic regions induce radial velocity perturbations through the Doppler rotation signal and the suppression of convective blueshift---collectively known as rotation--modulation. 
We developed the Rotation--Convection (RC) model: a method of detrending and characterizing rotation--modulation, using only cross--correlation functions or 1--dimensional spectra, without the need for continuous high cadence measurements. 
The RC method uses a simple model for the anomalous radial velocity induced by an active region and has two inputs: stellar flux (or a flux proxy) and the relative radial velocity between strongly and weakly absorbed wavelengths (analogous to the bisector--inverse slope).
On NEID solar data (three~month baseline), the RC model lowers the amplitude of rotationally--modulated stellar activity to below the meter--per--second level. 
For the standard star HD~26965, the RC model detrends the activity signal to the meter--per--second level for HARPS, EXPRES, and NEID observations, even though the temporal density and timespan of the observations differs by an order of magnitude between the three datasets.
In addition to detrending, the RC model also characterizes the rotation--modulation signal.
From comparison with the Solar Dynamics Observatory, we confirmed that the model accurately recovers and separates the rotation and convection radial velocity components.
We also mapped the amplitude of the rotation and convection perturbations as a function of height within the stellar atmosphere.
Probing stellar atmospheres with our revised spot model will fuel future innovations in stellar activity mitigation, enabling robust exoplanet detection.
\end{abstract}


\section{Introduction}
\label{sec:intro}
The radial velocity (RV) method has proved invaluable in the discovery and characterization of extrasolar planets \citep{Mayor1995, Butler1996, Mayor2011}.
In addition to white noise from photon-limited uncertainties, the noise floor for RV measurements is set by instrumental systematic errors,  telluric contamination, and the stellar variability of the host stars \citep[see][]{Isaacson2010,Luhn2020a}.
The current and next generation of extreme-precision spectrographs are lowering the instrumental systematics noise floor to at or well below the meter--per--second level: HARPS \citep{Mayor2003}, HARPS-N \citep{Cosentino2012}, 
ESPRESSO \citep{Megevand2014}, 
KPF~\citep{Gibson2016}, 
EXPRES \citep{Jurgenson2016}, 
NEID \citep{Schwab2016},
and Maroon-X \citep{Seifahrt2018}.
The discovery and characterization of extrasolar planets using RV measurements is now primarily limited by noise sources from stellar variability.

Multiple physical processes contribute to stellar variability.
From the shortest to longest variation timescales, the relevant activity mechanisms are:
(i) expansion and contraction of the stellar surface from \textit{pressure-mode} (p-mode) oscillations \citep[period of $5$~minutes and amplitude of order $\sim$$0.5$~{\ms} for the Sun,][]{Strassmeier2018}, (ii) changing \textit{granulation} patterns 
\citep[periods of minutes to days and amplitudes on the {\ms} level for the Sun,][]{Rieutord2010}, (iii) \textit{rotation--modulation} as spots and their associated magnetic regions move across the stellar disk  \citep[period of $25$ days and amplitude at the {\ms} level for the Sun,][]{AlMoulla2023}, and (iv) \textit{long-term activity} from changes in the spot-fraction on multi-year timescales. 
See the introduction section of \cite{deBeurs2020} for an in-depth review. 

Significant effort has gone into understanding and mitigating each of these anomalous RV signals.
The signatures of p--modes and granulation can be approximately averaged-out with longer exposure times and multiple observations per night \citep{Dumusque2011}. 
Targeted observing strategies can reduce the residual p-mode variations to $10$~{\cms} for the Sun \citep{Chaplin2019}; however, granulation induces RV variations on timescales ranging from hours to days \citep{Rincon2018}, limiting the effectiveness of averaging.
With a characteristic timescale of the stellar rotation period, rotation--modulation cannot be easily mitigated by observing strategy alone.
Rotation--modulation originates from the RV perturbations induced by spots and their associated magnetic regions. 
Spots and magnetic regions generate RV perturbations through two effects: (i) the darkness (brightness) of the spot (faculae) relative to the unspotted photosphere induces a flux imbalance between the rotationally red- and blueshifted hemispheres and (ii) the magnetic region locally suppresses convective blueshift \citep[see][]{Aigrain2012}.

Rotation--modulation is a significant barrier to planet detection and characterization with RVs. 
The RV perturbations are typically larger than the systematic noise floor for current and next generation spectrographs, and the variations can be mistaken for Keplerian due to their quasi-periodic nature.
Past data driven attempts to mitigate rotation--modulation include linear-regression to an activity indicator (e.g., strength of the Ca II lines, width of the cross--correlation function, or broadband photometry), Gaussian process regression \citep[e.g.,][]{Haywood2014,Dai2017,jones2022}, spectral line decomposition \citep[e.g.,][]{CollierCameron2021,Zhao2022FIESTA}, machine learning \citep[e.g.,][]{deBeurs2020,Liang2024}, and optimized observation scheduling \citep{Gupta2023}.
While the landscape of activity mitigation has grown rapidly, anomalous RV signals from stellar variability remain a barrier to exoplanet detection and characterization.
\cite{Zhao2022} applied 22 different methods to EXPRES observations of four low--to--moderate activity stars and found no method consistently reduced the activity perturbations to below the meter--per--second level.
Activity mitigation methods can also suffer from over-fitting \citep[e.g.,][]{Blunt2023}.

Physically motivated activity mitigation offers an alternative to the data driven methods enumerated above.
Models that leverage our knowledge of stellar activity can be less flexible, less vulnerable to overfitting, and more readily interpretable.
The simplified picture of rotation--modulation---i.e., the anomalous RV signal arises from the flux imbalance between the rotationally red- and blueshifted hemispheres and the suppression of convective blueshift---inspired the FF$^\prime$ method  \citep{Aigrain2012};
this method uses the simplified spot model to express the anomalous RV signal induced by an active region in terms of the host star's flux and the first time derivative of flux.
The FF$^\prime$ method is currently limited to datasets where photometry---or a flux proxy \cite[e.g.,][]{Rajpaul2015,Giguere2016}---is available and the temporal sampling is great enough for calculation of the first time derivative of flux. 

The motivation for the work described in this paper was to (i) increase the applicability of the spot model---i.e., relax the FF$^\prime$ method's requirement on high cadence observations---and (ii) explore the physics of stellar activity.
Investigations of stellar activity often rely upon spectrum averaged RV measurements.
For example, cross--correlation of the observed spectrum against a template of rest-frame spectral lines yields the cross--correlation function (CCF), i.e., the spectrum averaged line profile \citep{Baranne1996}.
The spectrum averaged (``bulk'') RV motion is then measured by locating the CCF's peak.
However, the rotation--modulation RV signal is known to be wavelength dependent.
Prior studies established that the rotation--modulation signal varies between shallow and deep spectral lines \citep{Cretignier2020} and between strongly and weakly absorbed wavelengths \citep{AlMoulla2022}.
These trends imply the rotation--modulation signal varies with altitude in the stellar atmosphere; wavelengths subjected to relatively high levels of absorption trace higher altitudes in the stellar atmosphere than wavelengths with comparatively lower levels of absorption \citep[][and see Section~\ref{sec:form_temp}]{Gray2009,AlMoulla2022}.
By considering how the rotation--modulation signal varies within the stellar atmosphere, we have developed a method of both detrending and \textit{characterizing} stellar activity induced RVs.

This paper is organized as follows. 
In Sections~\ref{sec:obs}~and~\ref{sec:extraction}, we describe the observations used in our study and our methods of extracting RV measurements, respectively.
We introduce our revised spot model in Section~\ref{sec:revisiting} and apply it in Section~\ref{sec:application}.
Our discussion and conclusions are presented in Sections~\ref{sec:discussion}~and~\ref{sec:conclusions}.

\section{Observations}
\label{sec:obs}

\begin{deluxetable*}{ccccccc | ccccc}[t]
\tablecaption{Summary of targets, instruments, and observations.\label{tab:obs}}
\tablehead{
 \colhead{Target} &  \colhead{M} & \colhead{T$_\mathrm{eff}$} &  \colhead{log $g$} & \colhead{[Fe/H]} & \colhead{$\log R^\prime_\mathrm{HK}$} &\colhead{Instrument} & 
 \colhead{N} & 
 \colhead{Start}  & \colhead{Baseline} &  \colhead{RMS}\\
 \colhead{} &  \colhead{(M$_\odot$)} &  \colhead{(K)} &  \colhead{} & \colhead{} & \colhead{} & \colhead{} & \colhead{} & \colhead{(BJD)}  & \colhead{(days)} &  \colhead{{\ms}}
}
\startdata
Sun & 1 & 5770 & 4.0 & 0.0 & $-4.91$ & NEID & 9081 & 2459636.2 & 98.0 & 2.05\\
\hline
{\HDtwo} & 0.76 & 5151 & 4.54 & $-$0.3 & $-4.99$ &HARPS & 248 & 2452939.8 & 1776.1 & 2.79\\
         & & & & & & EXPRES & 176 & 2458716.0 & 908.6 & 2.75\\
         & & & & & & NEID & 63 & 2459504.0 & 147.6 & 2.43\\
\enddata
\tablecomments{For both stars, we adopt $v_\mathrm{mic}=0.85$~{\kms} and $v_\mathrm{mac}=3.98$~{\kms} \citep{Valenti2005}. 
The properties of {\HDtwo} are from \cite{Diaz2018}.
The $\log R^\prime_\mathrm{HK}$ activity indices are quoted from \cite{Mamajek2008} and \cite{Jenkins2011} for the Sun and {\HDtwo}, respectively.
The number of observations is post--filtering (see Section~\ref{sec:obs}).
The reported RMS are from the spectrum averaged line by line RVs;
for the solar data, the RMS of the spectrum averaged CCF RVs is $2.21$~{\ms}.
}
\end{deluxetable*}

We studied two stars: the Sun and the well-studied RV standard star {\HDtwo}.
Solar observations have high signal-to-noise ratios (SNR) and enable comparison between disk-integrated RVs and spatially resolved images of the Sun.
{\HDtwo} is part of the NEID Earth Twin Survey \citep[NETS,][]{Gupta2021} and has been observed by several spectrographs over the last few decades.
As a moderately active star with an extensive history of RV observations, {\HDtwo} is a valuable testbed for activity mitigation strategies due to its inherent brightness ($V\sim4.4$) and clear rotation--modulation \citep{Ma2018,Diaz2018,Zhao2022,Laliotis2023,Burrows2024}.
Stellar parameters and selected observations are summarized in Table~\ref{tab:obs}.

For the Sun, we investigated three months of NEID observations from February 25th to June 3rd of 2022. 
A wildfire in Arizona in the summer of 2022 halted observatory operations, and required an emergency shutdown of the NEID spectrometer. 
Routine operations began again in November 2022.
Of observations prior to the fire and instrument shutdown, this time-span offers the longest continuous baseline with a detectable rotation--modulation RV signal.
To limit the effects of differential extinction \citep[the spurious RV signal induced by the extinction gradient over the solar disk,][]{Deming1987,CollierCameron2021}, we restricted ourselves to observations made within $1.5$~hours of solar-noon, i.e., a 3-hour wide window.
Each observation had an exposure time of 55 seconds and a readout time of 28 seconds, resulting in a cadence of 83 seconds.
The median pixel-level SNR was $415$. 
To discard observations with biased RVs---e.g., from clouds or instrument systematics---spectra were rejected if SNR~$<350$ and filtered by 4$\sigma$ clipping of the data reduction pipeline (DRP) RVs against a sinusoid plus a linear trend; the model was constructed to trace the rotation--modulation RV signal and constrained via regression to the DRP RVs, yielding a best-fit sinusoid period of $24.3$~days (consistent with the Sun's equatorial rotation-period).
Days with~$<30$ retained observations were discarded.
After all filters, 84\% of observations were retained (9081 observations from 81 days). 
Prior to RV extraction, the solar spectra were binned by day, effectively removing the impact of p--modes and averaging down granulation noise on multi--hour timescales.
Throughout this study, we used the binned solar spectra, rather than the individual spectra.

To contextualize the NEID solar data, we also used observations from the Helioseismic and Magnetic Imager (HMI) aboard the Solar Dynamics Observatory (SDO).
HMI observes the Sun near the magnetically sensitive $6173.3$~\AA ~FeI line at high spatial resolution \citep{Pesnell2012,Scherrer2012}.
Maps of continuum intensity, line-of-sight longitudinal magnetic field strength, and radial velocity are publicly available via \texttt{SunPy} \citep{SunPy2020}.
We used \texttt{solaster} to convert the spatially resolved HMI data into ``Sun-as-a-star" disk-integrated RVs \citep{Haywood2016,Ervin2022}.
\texttt{Solaster} calculates the rotation--modulation RV signal expected from the HMI maps of limb-darkening corrected intensity and magnetic field strength. 
The \texttt{solaster} RVs were calibrated to the NEID bandpass via linear regression \citep{Ervin2022}.
In Section~\ref{sec:phys}, we use the \texttt{solaster} ``Sun-as-a-star" RVs to validate the performance of our rotation--modulation detrending model.

For {\HDtwo}, we considered HARPS, EXPRES, and NEID observations.
Of the publicly available HARPS observations,\footnote{\url{https://dace.unige.ch}} we selected all observations made prior to May 2009---i.e., the longest continuous baseline prior to the fiber change on the 1st of June 2015.
We considered all EXPRES and NEID observations made prior to June 2022.
NEID data were collected as part of the NEID NETS program \citep{Gupta2021}. 
The observations were filtered for airmass~$<1.75$.
After the filter, 248/307, 176/187, and 63/63 observations were retained from HARPS, EXPRES, and NEID, respectively. 
The {\HDtwo} spectra were not binned.

To measure the RV from a given observation, we primarily relied on the 1--dimensional spectra;
for validation, spectrum averaged CCFs were also considered for the NEID solar data.
The NEID data-products were derived from the standard NEID DRP (version 1.2).\footnote{\url{https://neid.ipac.caltech.edu/docs/NEID-DRP/}}
The spectrum averaged CCFs were computed by combining the order-by-order CCFs provided in the NEID L2 files; the CCFs were calculated using the ESPRESSO G2 mask.\footnote{\url{https://www.eso.org/sci/software/pipelines/espresso/espresso-pipe-recipes.html}}
Individual order CCFs were co-added using the weights specified in the NEID L2 FITS header (\texttt{CCFWTNNN} in the \texttt{\lq{}CCFS\rq{}} extension, where NNN is the diffraction order number).
For EXPRES, the 1--dimensional spectra were produced by the EXPRES DRP \citep{Petersburg2020}.
These reductions include a polynomial-based wavelength solution and a non-parametric, hierarchical wavelength calibration generated using \texttt{excalibur} \citep{Zhao2021}. 
We adopted the \texttt{excalibur} wavelengths, which cover a narrower spectral range (approximately $5000$ to $7000$~\AA, the wavelengths covered by the EXPRES laser frequency comb) but show less instrumental scatter. 
For HARPS, we adopted the 1--dimensional spectra produced by the DRP pipeline.\footnote{\url{https://www.eso.org/sci/facilities/lasilla/instruments/harps/doc/DRS.pdf}}

\section{Radial Velocity Extraction}
\label{sec:extraction}

We used three methods for extracting RVs from stellar spectra: (i) line by line template matching, (ii) pixel by pixel template matching, and (iii) cross--correlation against a dictionary of spectral lines (i.e., CCFs).
As described below, the line by line method was used to measure spectrum averaged (``bulk'') RVs; the pixel by pixel method was used to calculate RVs for different ``slices'' of a spectrum (i.e., all pixels within a range of continuum normalized depths); the cross--correlation method was used as a soundness check and to ensure generality.

\subsection{Line by line}

Line by line RVs were measured using the pipeline of \cite{Siegel2022}; \cite{Dumusque2018} introduced the line by line method.
The line by line pipeline consists of three steps: (i) construct a reference spectrum; (ii) generate a spectral line list from the reference spectrum; (iii) perform template matching against the reference spectrum for each spectral line in each spectrum.
The line by line method was used to calculate the spectrum averaged RVs, instead of the different instruments' DRPs, to ensure the datasets were processed with a single pipeline.

For each combination of target and instrument, the reference spectrum was constructed by co-adding all spectra in the dataset (shifted into a common rest-frame).
The line list was generated from the reference spectrum following Appendix A of \cite{Cretignier2020}, as described briefly below. 
Lines were first identified in terms of local minima and maxima in the reference spectrum.
For two neighboring local maxima at $\lambda_{i, \rm left}$ and $\lambda_{i, \rm right}$, with a local minima between them at $\lambda_{i}$, we adopted a window width of $\min(\lambda_i-\lambda_{i, \rm left},\lambda_{i,\rm right}-\lambda_i)$ and a line center of $\lambda_{i}$.
The lines were then filtered for a minimum width of 10~pixels, a minimum depth relative to the local continuum of 5\%, and a separation of at least 48 pixels from telluric lines \citep[relative to a sample \texttt{TAPAS} sky spectrum,][]{Bertaux2014}.
Spectral lines were also filtered for line symmetry \citep[see Table C.1. of][]{Cretignier2020}.

For the $i$th line of the $j$th observation, the radial velocity RV$_{i,j}$ was measured by template matching. Following Eqn. 2 of \cite{Bouchy2001}, 
\begin{equation}
    S_{i,j}(\lambda) = A \Big[ S_{i,\rm reference}(\lambda) + \frac{\partial S_{i,\rm reference}(\lambda)}{\partial \lambda} \delta \lambda \Big],
\end{equation}
where $S_{i,j}(\lambda)$ is the $j$th observed spectrum within the $i$th line window, $S_{i,\rm reference}(\lambda)$ is the reference spectrum over the same interval, $A$ is the scale-factor offset between the observed and reference spectra, and $\delta \lambda$ is the wavelength offset between the two spectra.
The free--parameters $A\delta\lambda$ and $A$ were numerically inferred via least-squares regression.
The wavelength shift $\delta \lambda$ was converted to a radial velocity via $\text{RV}_{i,j}=\frac{c}{\lambda}\delta\lambda$;
the uncertainties in $A\delta\lambda$ and $A$ from the least-squares fit were propagated to calculate the RV uncertainty $\sigma_{\mathrm{RV},i,j}$.

As a low order proxy for changes in line shape, we also extracted the depth (relative to the local continuum) of each spectral line for each observation.
For the $i$th line of the $j$th spectrum, the line's depth is defined as
\begin{equation}
    \label{eqn:line_depth}
    d_{i,j} = 1 - \min[S_{i,j}(\lambda)] / C_{i,j},
\end{equation}
where $\min[S_{i,j}(\lambda)]$ is the minimum flux within the line window for the $j$th spectrum (calculated via cubic spline interpolation) and $C_{i,j}$ is the local continuum for the $i$th line (the average of the leftmost and rightmost pixel fluxes in the line’s window). 
For the Sun and {\HDtwo}, approximately $50\%$ of spectral lines are shallower than $0.5$, and $10\%$ of spectral lines are deeper than $0.9$.

After RV extraction, spectral lines were filtered via 4$\sigma$ clipping on RV$_{i,j}$, $\sigma_{\mathrm{RV},i,j}$, $d_{i,j}$, and $\sigma_{\mathrm{depth},i,j}$ \citep[motivated by][]{Dumusque2018}. 
For lines found in multiple spectral orders, the spectral order where the line had the lowest mean RV uncertainty was considered.
After applying all the filters described above, the solar dataset included approximately $3700$ spectral lines.
The {\HDtwo} datasets included approximately $2700$, $1300$, and $4000$ lines for HARPS, EXPRES, and NEID, respectively;
the lower number of spectral lines in the EXPRES dataset is consistent with the narrower spectral range of the \texttt{excalibur} wavelength solutions. 

\begin{figure*}[t]
\gridline{\fig{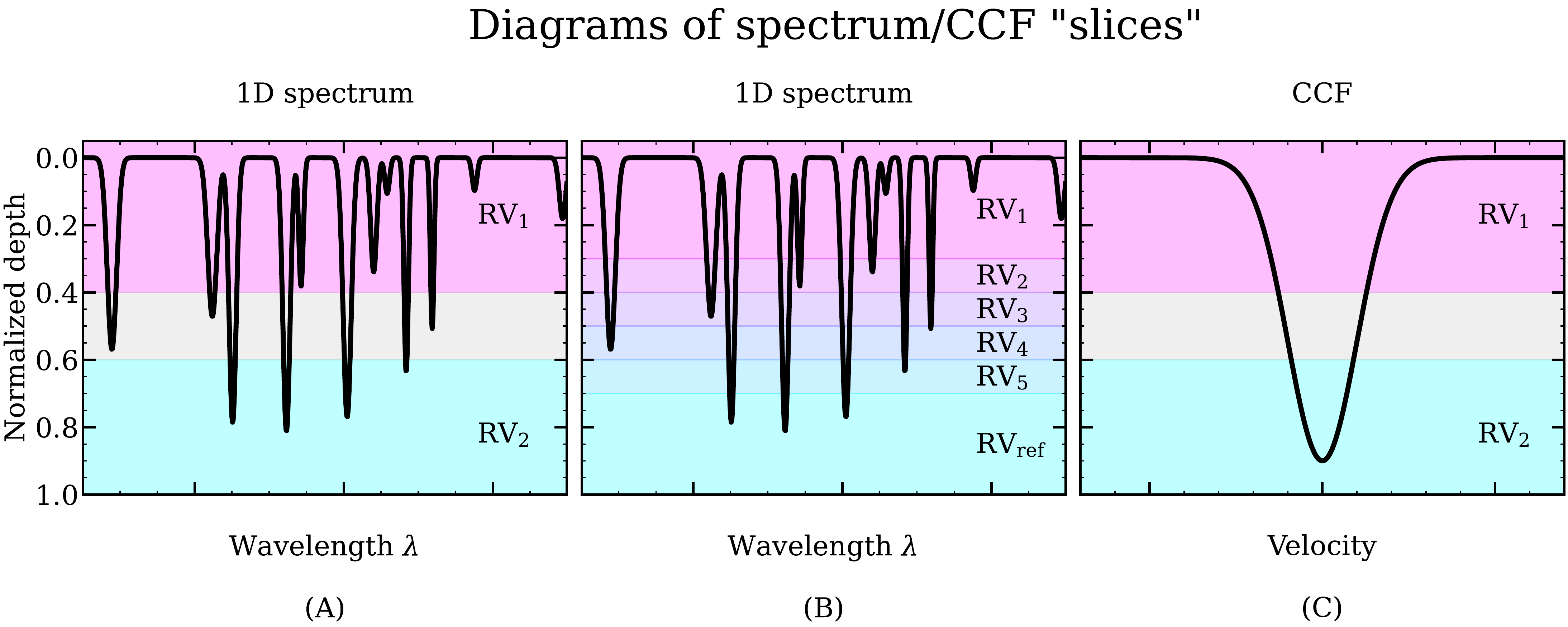}{
\textwidth}{} }
\caption{Throughout this study, we measured RVs from different ``slices'' of a 1--dimensional spectrum (or CCF).
For 1--dimensional spectra, we either considered two large slices or six narrow slices (leftmost and central panels).
We only considered the simpler case of two large slices (rightmost panel) for CCFs.
Regions not included in a slice are shaded grey.
}
\label{fig:slices_cartoon}
\end{figure*}

For a given spectrum, the line by line RV measurements were combined into a spectrum averaged RV via a weighted average (with weights $1/\sigma_{\mathrm{RV},i,j}^2$).
For the solar dataset, the average bulk RV uncertainty was only $2$~{\cms}; however, the solar spectra were daily binned prior to RV extraction, and the quoted RV uncertainty only reflects the propagated uncertainties from the template-matching process, which neglects daily RV scatter induced by instrumental systematics or intrinsic variability in the stellar spectrum (e.g., granulation).
For the {\HDtwo} datasets, the mean bulk RV uncertainties---again only photon-limited---for HARPS, EXPRES, and NEID were 52, 37, and 16 {\cms}, respectively.

The above procedure considers each dataset independently.
This is advantageous given the different optical-paths of HARPS, EXPRES, and NEID, e.g., a spectral line near the center of a spectral order in one instrument may fall near the edge of a spectral order in another.

\subsection{Pixel by pixel}
In addition to measuring spectrum averaged RVs, we also measured RVs for different ``slices'' of the 1--dimensional spectrum.
The slices were defined as ranges of continuum normalized depths (i.e., absorption strength);
this procedure is analogous to the measurement of a CCF's bisector.
Throughout this study, we either divided the spectrum into two large slices or six narrow slices (diagrammed in Figure~\ref{fig:slices_cartoon}).

Within the $i$th line window, the continuum normalized depth at the wavelength $\lambda$ is defined 
\begin{equation}
    \label{eqn:depth}
    d_{i}(\lambda) = 1 - S_{i,\mathrm{reference}}(\lambda) / C_{i,\mathrm{reference}}
\end{equation}{}
where $S_{i,\mathrm{reference}}(\lambda)$ is the reference spectrum flux at $\lambda$ and $C_{i,\mathrm{reference}}$ is the local continuum flux for the $i$th line. 
Unlike Eqn.~\ref{eqn:line_depth}, which defines the depth of a spectral line, Eqn.~\ref{eqn:depth} applies to individual pixels within the line window.
The pixel-level depths  $d_{i}(\lambda)$ were calculated from the reference spectrum, i.e., they do not vary from observation to observation.
To visualize the definition of $d_{i}(\lambda)$, Figure~\ref{fig:slices_cartoon} plots continuum normalized depth as a function of wavelength for a mock spectrum.

To measure the RV for a given spectral slice, we employed pixel by pixel RVs.
While template matching nominally has two free--parameters, the method can be extended to individual pixels if $A$ (the scale factor term) is held fixed---e.g., from a fit to the entire spectral line---and $\delta \lambda$ is solved for independently. 
For each pixel within an individual spectral line, $\delta \lambda$ and $ \sigma_{ \delta \lambda }$ were calculated by fixing $A$ and $\sigma_{A}$ to the best fit values derived from fitting the entire line. 
The RV of a slice was then measured by co-adding the pixel by pixel RVs for all pixels within the given range of depths.

\subsection{Cross correlation}
Lastly, we measured RVs from spectrum averaged CCFs. 
We used CCFs to validate the line by line and pixel by pixel methods;
CCFs are also a common data product of many instruments' DRPs, so considering CCFs ensures the methods presented below are widely applicable. 
For simplicity, we only considered CCFs for the NEID solar data.

As described in Section~\ref{sec:obs}, we adopted spectrum averaged CCFs from the NEID DRP.
For a given observation, we measured the RV via a least-squares regression between the spectrum averaged CCF and a Gaussian plus a linear trend.
We also measured the RV for the top $40\%$ and bottom $40\%$ of each CCF (analogous to bisector inverse slope); diagrammed in Figure~\ref{fig:slices_cartoon}.

Unless otherwise stated, the spectrum averaged RVs were measured from the line by line method and the RVs for spectral slices were measured from the pixel by pixel method.
The CCF derived RVs were only used for validation.
The line by line and CCF RVs are compared in Figure~\ref{fig:solar_summary} (for the NEID solar data).

\section{Revisiting the simplified spot model}
\label{sec:revisiting}

A spot and its associated magnetic region induce RV perturbations through two effects: (i) introduction of a flux imbalance between the rotationally red- and blueshifted hemispheres of the stellar disk and (ii) suppression of convective blueshift within the magnetized area.
Assuming the active region is small relative to the star's radius and neglecting differential rotation and limb darkening, the anomalous RV signature can be represented by two simple equations; see \cite{Aigrain2012} for a complete derivation.

The rotation perturbation is
\begin{align}
    \Delta \mathrm{RV}_\mathrm{rot}(t) &= -F(t) V_\mathrm{eq} \cos \delta \sin \phi (t) \sin i,
\end{align}
where $F(t)$ is the fractional change in a star's flux due to an active region located at latitude $\delta$ and longitude $\phi(t)$, $V_\mathrm{eq}$ is the star's equatorial rotation velocity, and $i$ is the star's inclination (the angle between the star's spin axis and the line of sight).
$F(t)$ depends on both the size and flux contrast (relative to the unspotted photosphere) of the active region, as well as its projected area. 
As the spot crosses the hemisphere facing the observer, $F(t)=f \cos \beta (t)$, where $f$ is the relative flux drop from an active region at the disk center and $\beta(t)$ is the angle between the line of sight and the spot normal.
For notational convenience, we refer to $F(t)$ as ``flux.''

The convection perturbation is analogously
\begin{align}
    \Delta \mathrm{RV}_\mathrm{C} (t) &= F(t) \delta V_c \kappa (\cos \phi(t) \cos \delta \sin i + \sin \delta \cos i),
\end{align}
where $ \delta V_c$ is the strength of convective blueshift suppression by the spot's associated magnetized region and $\kappa$ is the ratio of the magnetized area to the spot's area;
the terms within the parenthesis equal $\cos \beta (t)$.

As demonstrated by \cite{Aigrain2012}, these perturbations can be conveniently expressed in terms of the star's flux and the first time derivative of flux as
\begin{align}
    \Delta \mathrm{RV}_\mathrm{rot}(t) &= -F(t)F^\prime (t)R_{\star}/f,\\
    \Delta \mathrm{RV}_\mathrm{C}(t) &=   F^2(t)\delta V_c \kappa/f,
\end{align}
where $R_{\star}$ is the star's radius; this is commonly known as the FF$^\prime$ model.
For notational convenience, the parameters describing the active region---e.g., $\delta V_c$, $\kappa$, and $f$---are written as constants.
In reality, active regions grow and shrink, causing these parameters to vary with time; as discussed in Section~\ref{sec:RC_model}, we approximate this variability with linear functions of time.

The FF$^\prime$ framework has enabled physically motivated detrending of the rotation--modulation signal. 
The challenges of obtaining photometry and calculating its first time derivative have spurred several alterations. 
\cite{Giguere2016} demonstrated that the H$\alpha$ activity indicator serves as a photometric proxy and introduced a Gaussian smoothing kernel to ease calculation of the flux time derivative;
this version of the model requires high enough observing cadence to accurately approximate the flux time derivative and includes an additional free--parameter.
\cite{Rajpaul2015} adopted a Gaussian process approach, where the Ca H\&K activity indicator was used as a flux proxy and $F^\prime(t)$ was treated as a linear combination of Ca H\&K and the bisector inverse slope;
this approach bypasses the need to approximate the first time derivative of flux but is computationally more expensive.

In this section, we revisit the simplified spot model and derive a method of both detrending and characterizing the rotation--modulation RV signal.

\subsection{Relative Atmospheric Height}
\label{sec:form_temp}

To motivate our proposed model, here we briefly review how the rotation--modulation signal varies within the stellar spectrum (e.g., dependencies on wavelength and absorption strength).

Spectrum averaged RVs are often desirable for investigating Keplerian signals; Doppler shifts affect all wavelengths together, and considering the whole spectral range maximizes signal-to-noise of the resultant Doppler measurements.
However, some spectral lines can be more sensitive to systematic instrumental errors \citep{Dumusque2015}, while others are more sensitive to stellar magnetic activity \citep{Thompson2017,Dumusque2018,Wise2018,Cretignier2020};
this variability motivated the line by line approach \citep{Dumusque2018}.
Further complicating matters, the stellar variability RV signal can vary within a single spectral line.  

Photons that originate from a range of heights throughout the stellar atmosphere contribute to the observed spectrum.
At a given wavelength, the distribution of formation heights underlying the emergent flux is characterized by the cumulative contribution function: 
\begin{align}
    \label{eqn:cumulative_contribution_fn}
    C(\tau_\lambda )\propto \int_0^{\tau_\lambda} d \tau_\lambda^\prime S(\tau_\lambda^\prime) e^{-\tau_\lambda^\prime},
\end{align}
where $S$ is the source function and $\tau_\lambda$ is the optical depth at wavelength $\lambda$ \citep{Gray2005,AlMoulla2022}.
Approximating the photosphere as a 1--dimensional plane-parallel atmosphere, we can safely assume temperature and optical depth $\tau$ monotonically decrease with height.
Wavelengths subjected to relatively high levels of absorption are thus more sensitive to higher altitudes in the stellar atmosphere than wavelengths with comparatively lower levels of absorption.
Absorption relative to the local continuum is then correlated with atmospheric height.
Wavelength dependent continuum opacities slightly complicate this picture;
in the optical band, continuum opacity increases with wavelength \citep[driven by bound-free absorption of the negative hydrogen ion,][]{Gray2005}.
However, absorption line opacities are considerably stronger than continuum opacities, and continuum opacity varies on wavelength scales significantly greater than typical line widths.
Absorption relative to the local continuum is therefore an effective proxy for atmospheric height \citep[see Figure~2 of][]{AlMoulla2022}.
It follows that two slices of the 1--dimensional spectrum (i.e., different ranges of continuum normalized depths, see Figure~\ref{fig:slices_cartoon} for examples) will probe different average heights in the stellar atmosphere; 
in Section~\ref{sec:vs_atmo}, we quantify the correlation between absorption strength and atmospheric height using spectral synthesis.

Below, we consider how the rotation--modulation signal varies with atmospheric height (i.e., between different slices of the 1--dimensional spectrum).

\subsection{Rotation--Convection Model}
\label{sec:RC_model}

Observations indicate the RV perturbations induced by an active region vary with atmospheric height \citep{Gray2009,Liebing2021,Ellwarth2023,Cretignier2020,AlMoulla2022}.
Under this assumption, RVs drawn from different heights in the stellar atmosphere will have different contributions from the rotation and convection perturbations.

For rotationally--modulated RVs, the spectrum averaged RV signal can be decomposed as
\begin{align}
    \mathrm{RV}(t) &= \Delta \mathrm{RV}_\mathrm{rot}(t) + \Delta \mathrm{RV}_\mathrm{C}(t) + \Delta \mathrm{RV}_\mathrm{misc}(t) + K(t), \nonumber 
\end{align}
where $\Delta \mathrm{RV}_\mathrm{rot}(t)$ and $\Delta \mathrm{RV}_\mathrm{C}(t)$ are the spectrum averaged rotation--modulation RV perturbations, $K(t)$ is the Keplerian signal (if any), and $\Delta \mathrm{RV}_\mathrm{misc}(t)$ encapsulates other sources of RV variability---e.g., p--modes and granulation. 

In this work, we assume rotation--modulation is the dominant source of stellar variability. 
This choice was motivated by \cite{Luhn2023}, in which Gaussian process covariance kernels were used to predict the RV variability induced by p--modes and granulation as a function of stellar type and survey design.
We evaluated the \cite{Luhn2023} kernels using the actual observation times of our datasets, assuming integration times of $3$~hours and $10$~minutes for the Sun and {\HDtwo}, respectively.
For the solar dataset, the covariance kernels predict jitter levels of $3.7$~{\cms} and $11.5$~{\cms}, for oscillation and granulation, respectively (these values do not include photon-limited RV uncertainties).
For the NEID {\HDtwo} dataset, the expected jitter levels are $24$~{\cms} and $21$~{\cms}, for oscillation and granulation, respectively.
Based on these estimates, we expect the rotation--modulation signal to dominate the solar and {\HDtwo} RVs.  
Simultaneously modeling the effects of granulation and rotation--modulation is an important subject for future work.

If rotation--modulation is the dominant source of anomalous RVs, the measured velocity at a given atmospheric height $\mathrm{RV}_i(t)$ is
\begin{equation}
    \mathrm{RV}_i(t) = A_i \Delta \mathrm{RV}_\mathrm{rot}(t) + B_i \Delta \mathrm{RV}_\mathrm{C}(t) + K(t),
\end{equation}
where the coefficients $A_i,B_i$ reflect the dependence of the rotation and convection perturbations on height.
In practice, we measured a given $\mathrm{RV}_i(t)$ from a slice of the 1--dimensional spectrum (i.e., using only pixels within a certain range of continuum normalized depths); as validation, we also measured the RVs from different slices of a CCF (analogous to the bisector).

If we consider two different heights in the stellar atmosphere, it follows:
\begin{align}
    \label{equ:RV1-RV2}
    \mathrm{RV}_1(t) - \mathrm{RV}_2(t) &=  A_{12} \Delta \mathrm{RV}_\mathrm{rot}(t) + B_{12} \Delta \mathrm{RV}_\mathrm{C}(t), 
\end{align}
where $A_{12}\equiv A_1-A_2$ and $B_{12}\equiv B_1-B_2$;
Keplerian motion affects all wavelengths uniformly, so the $ K(t)$ terms cancel.

In terms of the simplified spot model, $\Delta \mathrm{RV}_{12} \equiv \mathrm{RV}_1(t) - \mathrm{RV}_2(t)$ is
\begin{align}
    \Delta \mathrm{RV}_{12}(t) &= A_{12} \left [  -F(t)F^\prime (t) \frac{R_{\star}}{f} \right] +  B_{12} \left [ F^2(t) \frac{\delta V_c \kappa}{f} \right].
\end{align}
The first time derivative of flux is then
\begin{align}
    F^\prime (t) &= - \left[ \Delta \mathrm{RV}_{12} -  B_{12} \frac{\delta V_c \kappa}{f} F^2(t)  \right] \frac{ f }{ A_{12} R_{\star} F(t) }.
\end{align}
Substituting the above equation into the simplified spot model, we see the rotation--modulation signal is
\begin{align}
    \label{eqn:deltaRVrot}
    \Delta \mathrm{RV}_\mathrm{rot}(t)&= \left[ \Delta \mathrm{RV}_{12} -  B_{12} \frac{\delta V_{c}\kappa}{f} F^2(t)  \right] \frac{ 1 }{ A_{12}},\\
    \label{eqn:deltaRVc}
    \Delta \mathrm{RV}_\mathrm{C}(t)&=  F^2(t)\delta V_c \kappa/f.
\end{align}
This formulation does not involve numerically taking the time derivative of stellar flux; the above model therefore does not require high-cadence observations, unlike the FF$^\prime$ method.
As explored below, this formulation also enables novel studies of the stellar atmosphere.
Given its generality, we refer to our revised spot model as the \textit{Rotation--Convection} (RC) model;
however, we stress that the general spot model underlying the RC model is identical to the FF$^\prime$ method of \cite{Aigrain2012}.  

The RC model has two inputs: (i) the relative RV between two different heights in the stellar atmosphere $\mathrm{RV}_{1}(t)-\mathrm{RV}_{2}(t)$ and (ii) stellar flux $F(t)$.

Heavily absorbed wavelengths are more sensitive to higher altitudes in the stellar atmosphere than wavelengths with comparatively lower levels of absorption. 
To compare the RVs of two different heights in the stellar atmosphere, we therefore measured the RVs of two different slices of the 1--dimensional spectrum (diagrammed in Figure~\ref{fig:slices_cartoon}); as a soundness check, we also measured the relative RV between the top and bottom of spectrum averaged CCFs (analogous to bisector inverse slope).

Complicating matters, the rotation component of rotation--modulation will vary within a spectral line independent of formation height.
For a rotating star, the observed stellar spectrum is, to first order, a convolution of the disk-integrated rest-frame spectrum with the rotation profile---the intensity-weighted distribution of velocities from the observed stellar surface. 
In the absence of any surface features, the rotation profile peaks at RV$=0$~{\ms} (ignoring any net convective blueshift) and monotonically falls with |{RV}| until it reaches $0$ at $\pm 2 \pi R_\star / P_\mathrm{rot}$ \citep[see Figure 1 of][]{DiMaio2023}.
A small spot (faculae) at latitude $\delta$ and longitude $\phi$ distorts the rotation profile by changing the relative contribution of $\mathrm{RV(\delta, \phi)}$ to the intensity-weighted RV distribution---e.g., a spot near the edge of the stellar disk will perturb the wings of lines, while a spot near the center of the disk will perturb the cores of lines. 
These perturbations will be approximately constant as a function of depth relative to the line core (modulo any variability in the relative spot contrast as a function of wavelength).
A given spectral slice (i.e., a range of depths relative to the continuum) will sample different parts of the rotational profile for each spectral line.
We therefore expect the strength of the rotation perturbation to vary between two spectral slices, regardless of whether $\Delta \mathrm{RV}_\mathrm{rot}(t)$ depends on atmospheric height.

This ambiguity does not hinder the performance of the RC model.
By construction, the RC model only assumes $\mathrm{RV}_{1}(t)$ and $\mathrm{RV}_{2}(t)$ have different contributions from $\Delta \mathrm{RV}_\mathrm{rot}(t)$ and $\Delta \mathrm{RV}_\mathrm{C}(t)$. The model is agnostic to whether those differences depend on formation height, depth in the line profile, or a combination of the two.

The RC model also requires a flux timeseries $F(t)$.
We approximated $F(t)$ by assuming a linear relationship between an activity indicator and flux \citep[e.g.,][]{Giguere2016}. 
Changes in line depth were used as the flux proxy.
We adopted the depth metric \citep[coadding the variations of many activity sensitive spectral lines,][]{Siegel2022} as our fiducial flux proxy; as a soundness check, we also considered CCF contrast (the difference between the CCF's peak and the pseudo--continuum).
Since the flux variations are small, we made the approximation: 
$ (F(t)-\langle F(t) \rangle)^2 \approx ( aP(t) + b)^2 \approx c P(t), $
where $P(t)$ is the flux proxy (normalized such that $\langle P(t) \rangle =0$) and $P^2(t) \ll P(t)$.
This simplification limits the number of free--parameters and does not noticeably change the results.
With the use of flux proxies, our revised spot model can be applied using only standard RV data-products: 1--dimensional spectra or CCFs.

In Section \ref{sec:application}, we explore the applications of the RC model, including (i) detrending the rotation--modulation signal, (ii) inferring the underlying $\Delta \mathrm{RV}_\mathrm{rot}(t)$ and $\Delta \mathrm{RV}_\mathrm{C}(t)$ perturbations, and (iii) mapping the strength of the rotation and convection perturbations as a function of relative height in the stellar atmosphere.

\section{Application}
\label{sec:application}

\subsection{Stellar activity detrending}
\label{sec:detrending}

\begin{figure*}[t]
\gridline{\fig{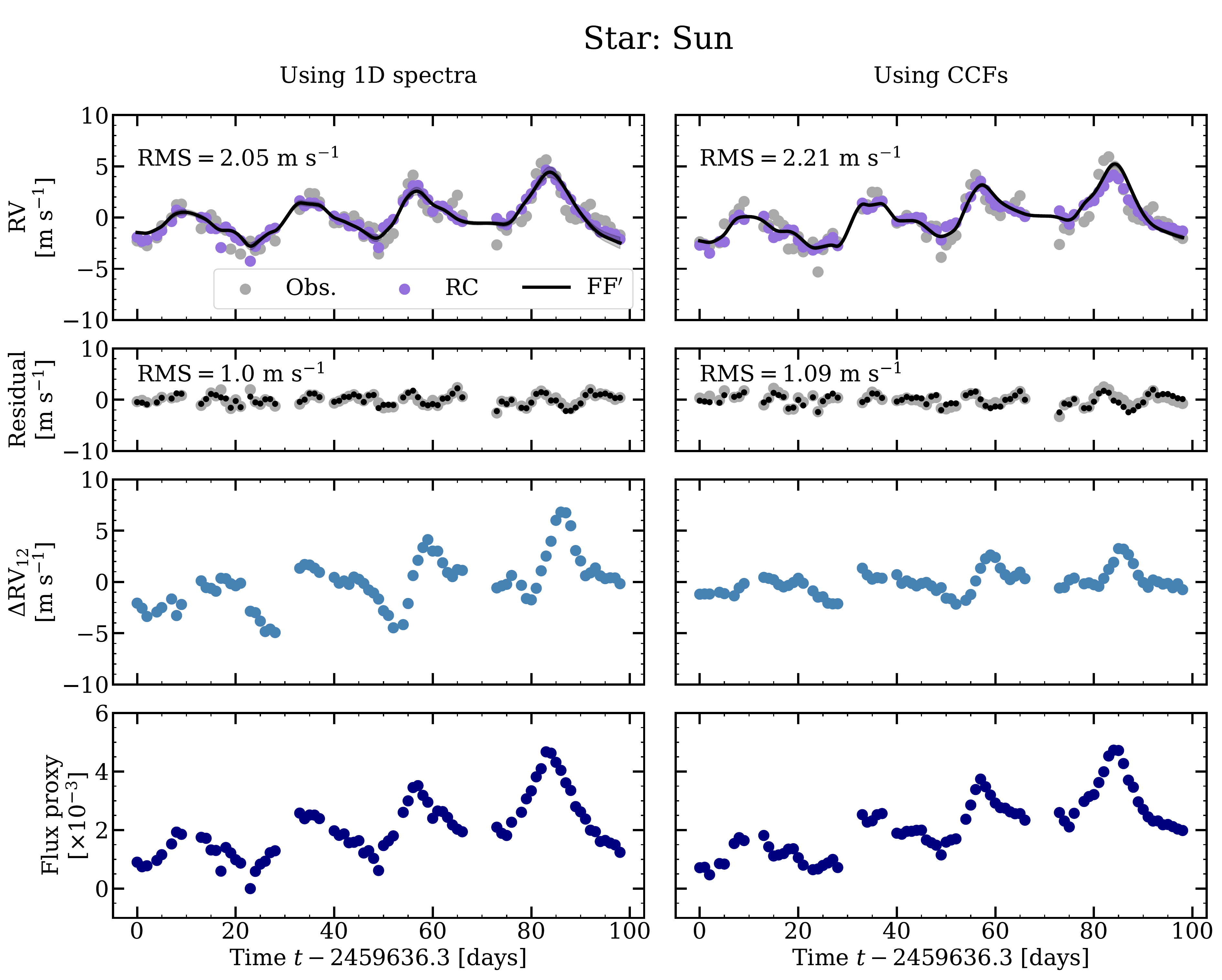}{
\textwidth}{} }
\caption{ The RC model successfully detrends the rotation--modulation activity signal in NEID solar RVs. 
The columns correspond to different methods of applying the model: either using the 1--dimensional spectra (left) or using only the CCFs (right).
Top row: spectrum averaged NEID solar RVs (grey) alongside the best fit RC detrending model (purple); the spectrum averaged RVs were measured either from the 1--dimensional spectra by the line by line method (left) or from the CCFs (right).
The RMS of the observed RVs is reported for each method. 
For comparison, FF$^\prime$ fits are shown in black; the 1$\sigma$ confidence interval is indistinguishable.
Second row: residuals between the spectrum averaged RVs and the best fit RC detrending model (grey);
the residuals between the observed RVs and the median FF$^\prime$ model are shown in black.
Third row: $\Delta$RV$_{12}$, the relative RV between two slices of the 1--dimensional spectrum (left column) or CCF (right column); the slices are shown in the left- and rightmost panels of Figure~\ref{fig:slices_cartoon}.
Bottom row: the flux proxy, either the depth metric (left) or CCF contrast (right).}
\label{fig:solar_summary}
\end{figure*}

\begin{figure*}[t]
\gridline{\fig{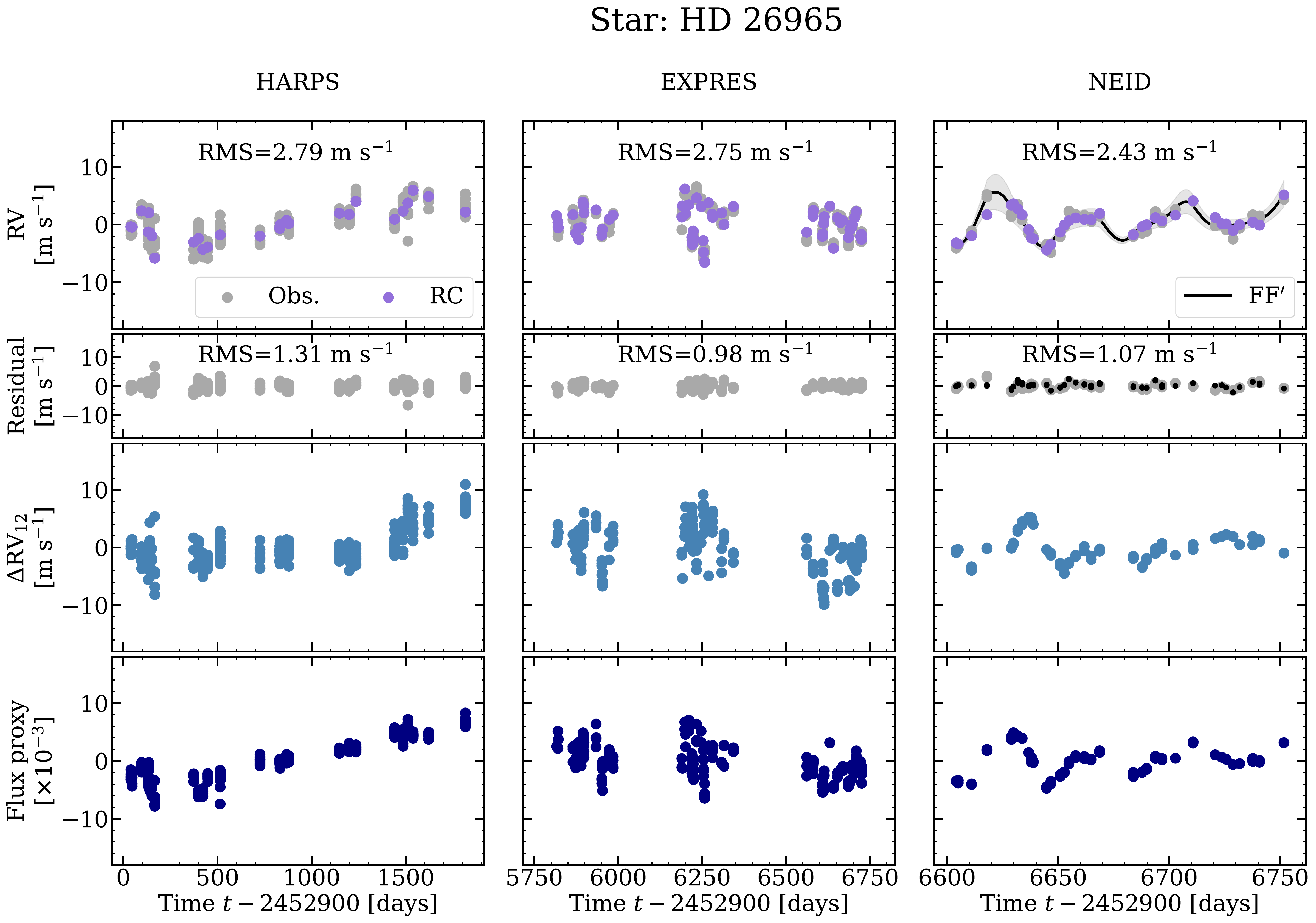}{
\textwidth}{} }
\caption{ Across the three {\HDtwo} datasets (each of which has a significantly different cadence), the RC model successfully traces the rotation--modulation signal. 
The NEID and EXPRES measurements offer high cadence coverage of multiple stellar rotation periods, while the HARPS data captures the multiyear activity cycle. 
Top row: spectrum averaged line by line RVs (grey) alongside the best fit RC detrending model (purple);
the RMS of the observed RVs is reported for each dataset.
For context, a FF$^\prime$ fit to the NEID data is shown in black; the shaded region demarcates the 1$\sigma$ confidence interval.
Second row: residuals between the observed RVs and the best fit RC detrending model (grey); the residuals between the observed RVs and the median FF$^\prime$ model are shown in black.
Third row: $\Delta$RV$_{12}$, the relative RV between two slices of the 1--dimensional spectrum (i.e., different ranges of continuum normalized depth); the slices are diagrammed in the leftmost panel of Figure~\ref{fig:slices_cartoon}.
Bottom row: depth metric activity indicator. 
}
\label{fig:rv_comparison}
\end{figure*}

\begin{figure*}[t]
\gridline{\fig{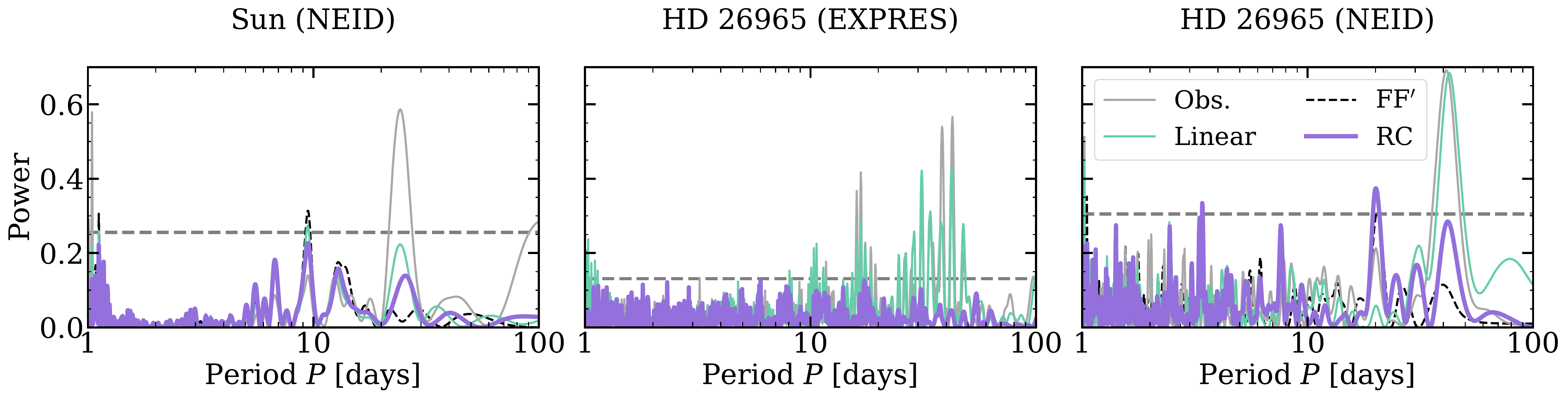}{
\textwidth}{} }
\caption{
The RC model greatly mitigates the stellar activity signal for the Sun and {\HDtwo}. 
We present Lomb--Scargle periodograms of the spectrum averaged RVs (grey), the RVs post-linear detrending with the flux proxy (green), and the RVs post-RC detrending (purple).
The NEID datasets also include FF$^\prime$ detrending (dashed) for reference.
Periodograms are shown for the NEID solar data, as well as the EXPRES and NEID {\HDtwo} data (the HARPS data is omitted due to the coarse sampling); the horizontal lines show the 1\% false--alarm--probabilities.
The presented results were all derived from the 1--dimensional spectra (not the CCFs).}
\label{fig:periodograms}
\end{figure*}
We first present the limiting case where the RC model is applied solely for stellar activity detrending. 
In terms of observables---$F(t)$ and $\Delta \mathrm{RV}_{12}(t)$---the RC model simplifies to
\begin{align}
    \Delta \mathrm{RV}_\mathrm{rot}(t) + \Delta \mathrm{RV}_\mathrm{C}(t)
    &=  \alpha_{12} F^2(t) + \beta_{12} \Delta \mathrm{RV}_{12}(t),
\end{align}
where $\alpha_{12}\equiv \delta V_c \kappa (1-B_{12}/A_{12})/f $ and $\beta_{12}\equiv A_{12}^{-1}$.
The above equation hides the physical meaning of the free--parameters and no longer separates $\Delta \mathrm{RV}_\mathrm{rot}(t)$ and $\Delta \mathrm{RV}_\mathrm{C}(t)$.
This simplification reduces the number of free--parameters and parameter--parameter degeneracies.
The detrending results with the simplified model are identical to the complete RC method.
At the cost of physical interpretability, the RC model can now be optimized using least-squares regression instead of more expensive MCMC sampling.

Complicating matters, active regions are not static phenomena.
We expect the properties of the active regions---e.g., the spot fraction $f$ or the strength of convective blueshift suppression $\delta V_c$---to vary with time.
Under the assumption that the active regions vary slowly (relative to the stellar rotation period),
we treated $\beta_{12}$ as a constant and $\alpha_{12}$ as a linear function of time.
To account for long-term changes in the properties of the active regions, we modeled each observing semester independently.
As discussed in Section~\ref{sec:discussion}, these assumptions could be relaxed by treating the free--parameters as Gaussian processes \citep[e.g.,][]{Rajpaul2015}.

We applied the simplified RC model to the Sun and {\HDtwo}; summarized in Figures~\ref{fig:solar_summary}~and~\ref{fig:rv_comparison}.
For both stars, the RC model accurately detrends the rotation--modulation signal.

For the NEID solar data, we tested the performance of the RC model using either the 1--dimensional spectra or the spectrum averaged CCFs.
In the first case, spectrum averaged RVs were calculated by the line by line method, the line by line depth metric was treated as the flux proxy, and $\Delta \mathrm{RV}_{12}(t)$ was measured from two large spectral slices (diagrammed in the leftmost panel of Figure~\ref{fig:slices_cartoon}).
As validation, we repeated the detrending process using the spectrum averaged CCFs.
Spectrum averaged RVs were measured from Gaussian fits to the CCFs, CCF contrast was adopted as the flux proxy, and $\Delta \mathrm{RV}_{12}$ was calculated using the top and bottom 40\% of the CCFs, respectively.
In both cases, the RC model successfully tracked the rotation--modulation signal, reducing the RMS from $2$ to $1$~{\ms}.
For each method, Figure~\ref{fig:solar_summary} presents the spectrum averaged RVs alongside the best-fit RC model; the ingredients of the model---the flux proxy and $\Delta \mathrm{RV}_{12}(t)$---are presented as well.
Differences between between the line by line and CCF methods are minimal and potentially originate from the different line lists used.

\begin{figure*}[t]
\gridline{\fig{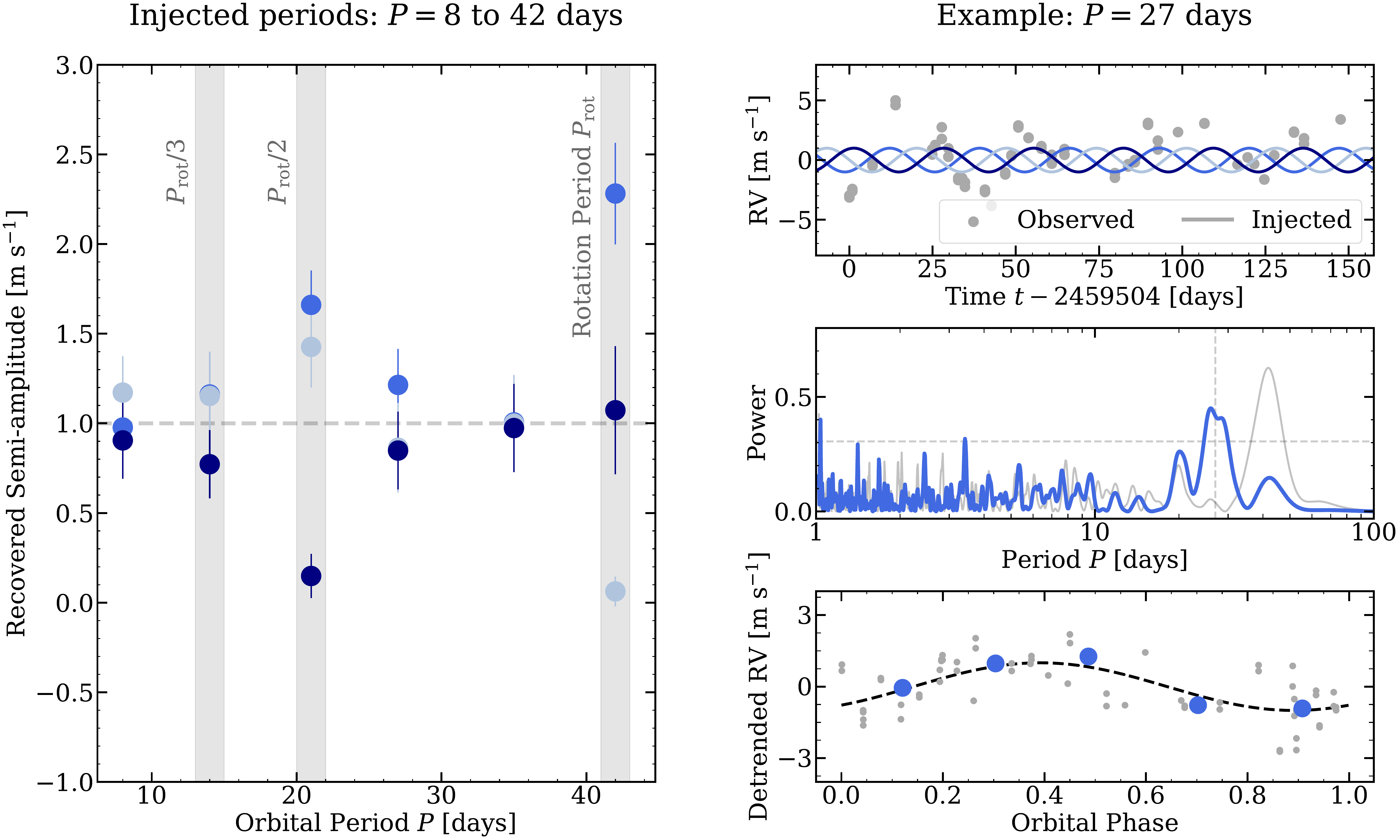}{
\textwidth}{} }
\caption{ 
Injected Keplerian signals were successfully recovered with the RC model. 
Using the NEID {\HDtwo} dataset, we considered a single planet on a circular orbit with a semi-amplitude of $K=1$~{\ms}.
Left: recovered semi-amplitudes for different combinations of the injected planet's orbital period and phase. 
The approximately $42$~day rotation period of {\HDtwo}, along with the lowest--order harmonics, are demarcated by shaded regions.
The colors correspond to the orbital phase of the injected planet (diagrammed in the upper right panel).
Right: breakdown of the injection--recovery process for $P=27$~days.
Upper row: spectrum averaged RVs (grey) alongside the injected Keplerian signal for three different orbital phases (described in Section~\ref{sec:keplerian}).
Middle row: periodograms of the RV timeseries before (grey) and after (blue) the activity component of the model was subtracted;
the vertical line demarcates the injected planet's orbital period and the horizontal line shows the 1\% false--alarm--probability.
Bottom row: activity detrended phase--folded RVs against the injected Keplerian signal (dashed line); phase--binned RVs are shown in blue. }
\label{fig:keplerian_recovery}
\end{figure*}

\begin{figure*}[t]
\gridline{\fig{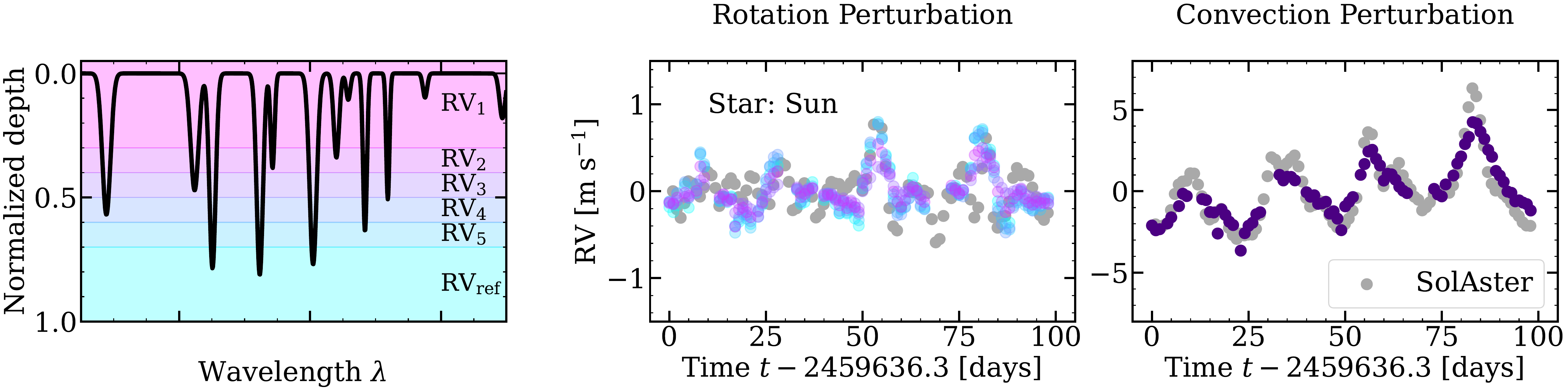}{
\textwidth}{} }
\caption{ The RC model recovers the rotation $\Delta \mathrm{RV}_\mathrm{rot}(t)$ and convection $\Delta \mathrm{RV}_\mathrm{C}(t)$ rotation--modulation perturbations from the NEID solar data, using only the 1--dimensional spectra.
To reign in parameter-parameter degeneracies, the RC model was jointly fit for five different choices of $\Delta \mathrm{RV}_{i,\mathrm{ref}}(t)=\mathrm{RV}_{i}(t)-\mathrm{RV}_{\mathrm{ref}}(t)$.
Left: diagram of how RV$_i$ and RV$_\mathrm{ref}$ were defined, in terms of continuum normalized depth (identical to the center panel of Figure~\ref{fig:slices_cartoon}). 
Center: the recovered timeseries of $\Delta \mathrm{RV}_\mathrm{rot}(t)$ from the RC model, compared to the independent SDO measurements (grey).
Each choice of $\Delta \mathrm{RV}_{i,\mathrm{ref}}(t)$ yields a prediction for $\Delta \mathrm{RV}_\mathrm{rot}(t)$;
all five predictions are presented, following the color scheme of the spectrum diagram.
Right: the recovered timeseries of $\Delta \mathrm{RV}_\mathrm{C}(t)$ compared to SDO;
the convection perturbation depends on the flux proxy alone, so there is only one prediction.
From marginalizing over the posteriors, the $1\sigma$ uncertainties on the model's rotation and convection components are $13$ and $9$~{\cms}, respectively.
}
\label{fig:RVspot_RVconv_recovery}
\end{figure*}

\begin{figure}[t]
\gridline{\fig{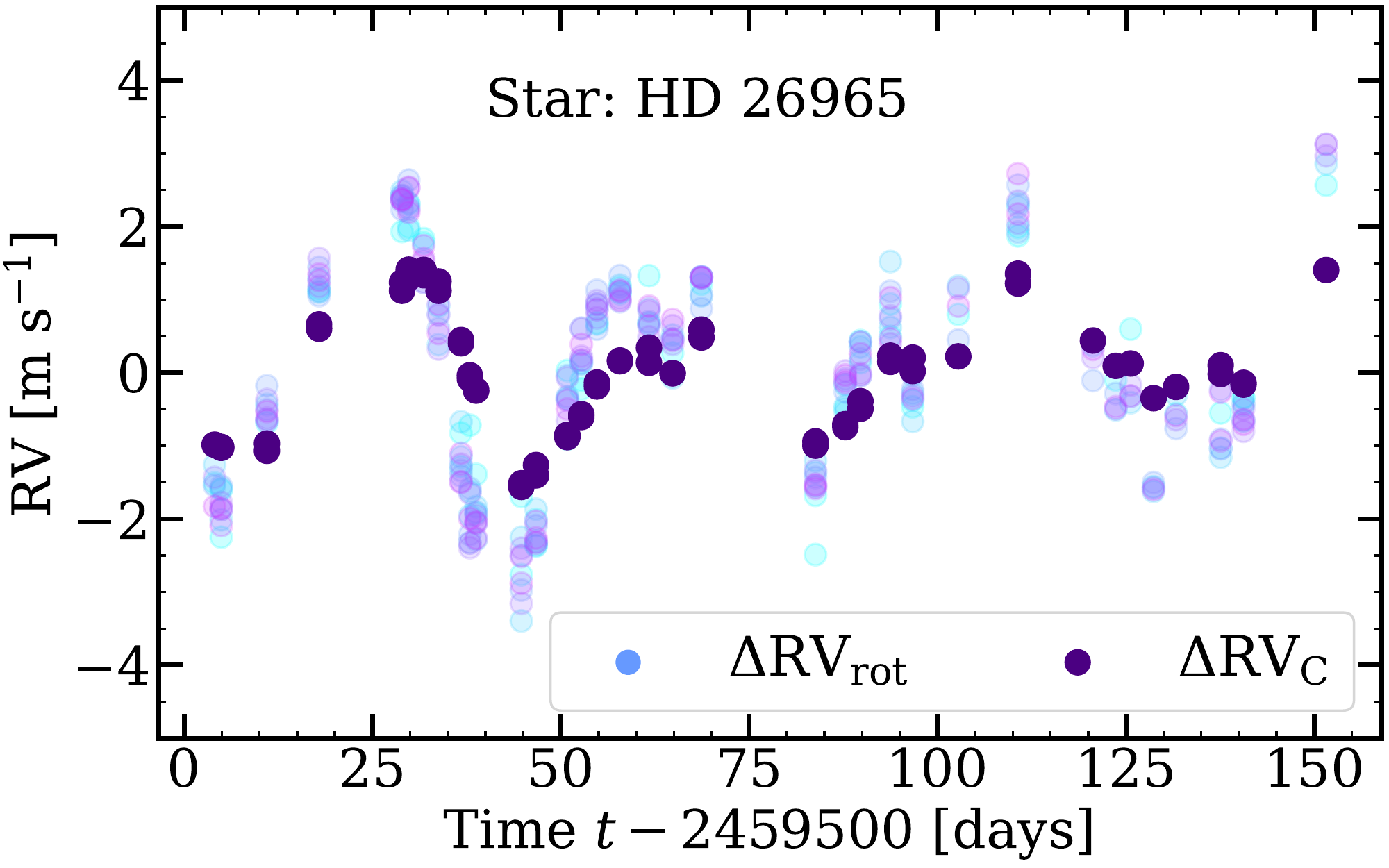}{
\columnwidth}{} }
\caption{
Inferred rotation $\Delta \mathrm{RV}_\mathrm{rot}(t)$ and convection $\Delta \mathrm{RV}_\mathrm{C}(t)$ perturbations for the NEID {\HDtwo} dataset, using only the 1--dimensional spectra.
To reign in parameter-parameter degeneracies, the RC model was jointly fit for five different choices of $\mathrm{RV}_{i,\mathrm{ref}}(t)=\mathrm{RV}_{i}(t)-\mathrm{RV}_{\mathrm{ref}}(t)$, diagrammed in Figure~\ref{fig:RVspot_RVconv_recovery}.
The recovered $\Delta \mathrm{RV}_\mathrm{rot}(t)$ signals (one for each choice of $\Delta \mathrm{RV}_{i,\mathrm{ref}}$) are presented following the color scheme of Figure~\ref{fig:RVspot_RVconv_recovery}.
The recovered timeseries of $\Delta \mathrm{RV}_\mathrm{C}(t)$ is presented in solid-purple.
From marginalizing over the posteriors, the $1\sigma$ uncertainties on the model's rotation and convection components are $15$ and $9$~{\cms}, respectively.
}
\label{fig:RVspot_RVconv_HD26965}
\end{figure}

For the {\HDtwo} datasets, the RC model reduced the RV RMS to near $1$~{\ms}.
Only the 1--dimensional spectra were considered for simplicity.
Figure~\ref{fig:rv_comparison} presents the spectrum averaged RVs for each instrument alongside the best-fit RC model; the inputs to the RC model are included for context.
While the residual RMS is similar across the three instruments, the observing cadence differs by over an order of magnitude; the median time between observations was 34, 7, and 3 days for HARPS, EXPRES, and NEID, respectively.
The similarity of the results, despite the range of observing cadences, is a key benefit of our revised spot model.

For comparison, we also performed ``standard'' FF$^\prime$ detrending---i.e., inferring the first time derivative of flux via numerical differentiation.
Directly following \cite{Giguere2016} and \cite{Siegel2022}, the photometric proxy was smoothed with a rolling Gaussian and the first time derivative was calculated through cubic-spline interpolation; the Gaussian's width was treated as a free parameter.
The model was optimized via MCMC sampling with \texttt{emcee} \citep{ForemanMackey2013};
uniform priors were adopted on the free--parameters.
Without densely sampled observations, the first time derivative of flux cannot be inferred numerically.
The FF$^\prime$ method is therefore only applicable to the solar observations and the NEID {\HDtwo} dataset.
For both the Sun and {\HDtwo}, the FF$^\prime$ method was applied to the spectrum averaged line by line RVs, with the depth metric as the flux proxy. 
As a soundness check (performed on the solar data), we also applied the FF$^\prime$ method to the spectrum averaged CCF RVs, with CCF contrast as the flux proxy.
The FF$^\prime$ models are presented in Figures~\ref{fig:solar_summary}~and~\ref{fig:rv_comparison}.

The RC and FF$^\prime$ models consistently agree.
For the Sun, the FF$^\prime$ model achieves residual RMS of $1.01$ and $1.09$~{\ms} using the 1--dimensional spectra and CCFs, respectively.
For {\HDtwo}, the FF$^\prime$ model achieves a residual RMS of $0.93$~{\ms}.
The residual structure in the detrended RVs is nearly identical between the RC and FF$^\prime$ approaches.
While the RC and FF$^\prime$ models detrend the rotation--modulation signal with the same quality, the RC method notably lacks the requirement of high cadence observations. 

The lingering periodic signals in the detrended RVs are summarized in the periodograms in Figure~\ref{fig:periodograms}.
In addition to RC and FF$^\prime$ modeling, the datasets were also detrended via linear regression against the flux proxy.
For each dataset, the RC model outperforms linear regression and closely matches the FF$^\prime$ results. 
Due to the sparse sampling, the HARPS RVs show negligible power in the periodograms and are not included for clarity.
The RC and FF$^\prime$ models significantly reduce the power at the stars' rotation periods (approximately 27 and 42 days for the Sun and {\HDtwo}, respectively).
For both stars, the periodograms reveal residual structure in the detrended RVs.
This structure likely originates from our reliance on a flux proxy.
A time lag between flux and the flux proxy and/or the relationship between flux and the flux proxy varying with time naturally results in structured residuals \citep[e.g.,][]{Burrows2024}; we discuss potential remedies in Section~\ref{sec:discussion}.

The RC model closely traces the rotation--modulation RV signal, with minimal requirements on observing cadence and computational resources. 
For both the Sun and standard star {\HDtwo}, the RC and FF$^\prime$ methods closely match.
Given its success describing the rotation--modulation signal, below we leverage the model's interpretability to better characterize stellar activity and explore stellar atmospheres.

\subsection{Keplerian injection--recovery}
\label{sec:keplerian}

To ensure the RC model preserves Keplerian information, we conducted a series of injection tests.
Using the NEID {\HDtwo} dataset, a planet was injected on a circular orbit with a semi-amplitude of $K=1$~{\ms}.
We considered six orbital periods between $8$ and $42$~days and three starting orbital phases: $(t_0-t_\mathrm{min})/P = 0.2, 0.5,$~and~$0.8$, where $t_\mathrm{min}$ is the time of the first observation and $P$ is the injected orbital period.
To recover the injected planet signal, the RVs were detrended with a joint RC and Keplerian model.
Figure~\ref{fig:keplerian_recovery} presents the posteriors on the recovered RV semi-amplitude for each combination of orbital period and phase. 
Over the 150~day baseline, {\HDtwo}'s rotation--modulation RV signal is quasi-sinusoidal, with a near constant amplitude.
Since the amplitude of the rotation--modulation signal is free to vary in the RC model, planets with orbital periods near the stellar rotation period (or its harmonics) are difficult to recover.
Since the amplitude and phase of the rotation--modulation signal changes with time (e.g., from changes in the spot distribution), the degeneracy between the Keplerian and stellar activity signals can be suppressed by considering longer baselines.
With the exception of orbital periods near the stellar rotation period (or its harmonics), the recovered semi-amplitudes are statistically consistent with the injected value.
These experiments confirm the RC model successfully preserves Keplerian information.

\subsection{Recovering the rotation and convection perturbations}
\label{sec:phys}

We next task the RC model with inferring the underlying components of rotation--modulation---the rotation $\Delta \mathrm{RV}_\mathrm{rot}$ and convection $\Delta \mathrm{RV}_\mathrm{C}$ perturbations---using only 1--dimensional stellar spectra.

The inputs to the RC model are stellar flux and the difference between the RVs drawn from higher in the stellar atmosphere and those probing lower in the atmosphere.
In practice, we adopt a flux proxy \citep[e.g., the depth metric, see][]{Siegel2022} and consider the difference between the RVs of two spectral slices (i.e., different ranges of continuum normalized depths, diagrammed in Figure~\ref{fig:slices_cartoon}).
The free--parameters of the RC model are $\delta V_c \kappa /f$, which describes the properties of the active regions, and $A_{12},B_{12}$, which characterize the relative contributions of $\Delta \mathrm{RV}_\mathrm{rot}(t)$ and $\Delta \mathrm{RV}_\mathrm{C}(t)$ in the RVs of the two spectral slices RV$_1(t)$, RV$_2(t)$;
the linear relationship between flux and the flux proxy adds a degree of freedom.
Unfortunately, $\delta V_c \kappa /f$ and $B_{12}$ are degenerate. 
In Section~\ref{sec:detrending}, this degeneracy was bypassed by simplifying the RC model to a linear combination of the observables.
To retain the physical meaning of the free--parameters, here we reign in the degeneracy.

Consider $N$ slices of the 1--dimensional spectrum.
The RV of each slice---RV$_1$,\dots, RV$_N$---will probe a different average height in the stellar atmosphere.
The RC model could be optimized using any $\Delta \mathrm{RV}_{ij}(t) \equiv \mathrm{RV}_i(t) - \mathrm{RV}_j(t)$.
If the strengths of the rotation and convection components vary with height in the stellar atmosphere, the free--parameters $A_{ij}$ and $B_{ij}$---i.e., the relative strengths of the rotation and convection components between $\mathrm{RV}_i(t)$ and $\mathrm{RV}_j(t)$---will depend on the particular choice of  $\Delta \mathrm{RV}_{ij}(t)$.
However, $\delta V_c \kappa /f$---the spectrum averaged properties of the active region(s)---by definition is the same for any choice of $\Delta \mathrm{RV}_{ij}(t)$.
The degeneracy between $\delta V_c \kappa /f$ and $B_{ij}$ is therefore lifted by jointly fitting the RC model for multiple choices of $\Delta \mathrm{RV}_{ij}(t)$, using a common $\delta V_c \kappa /f$.

\begin{figure*}[t]
\gridline{\fig{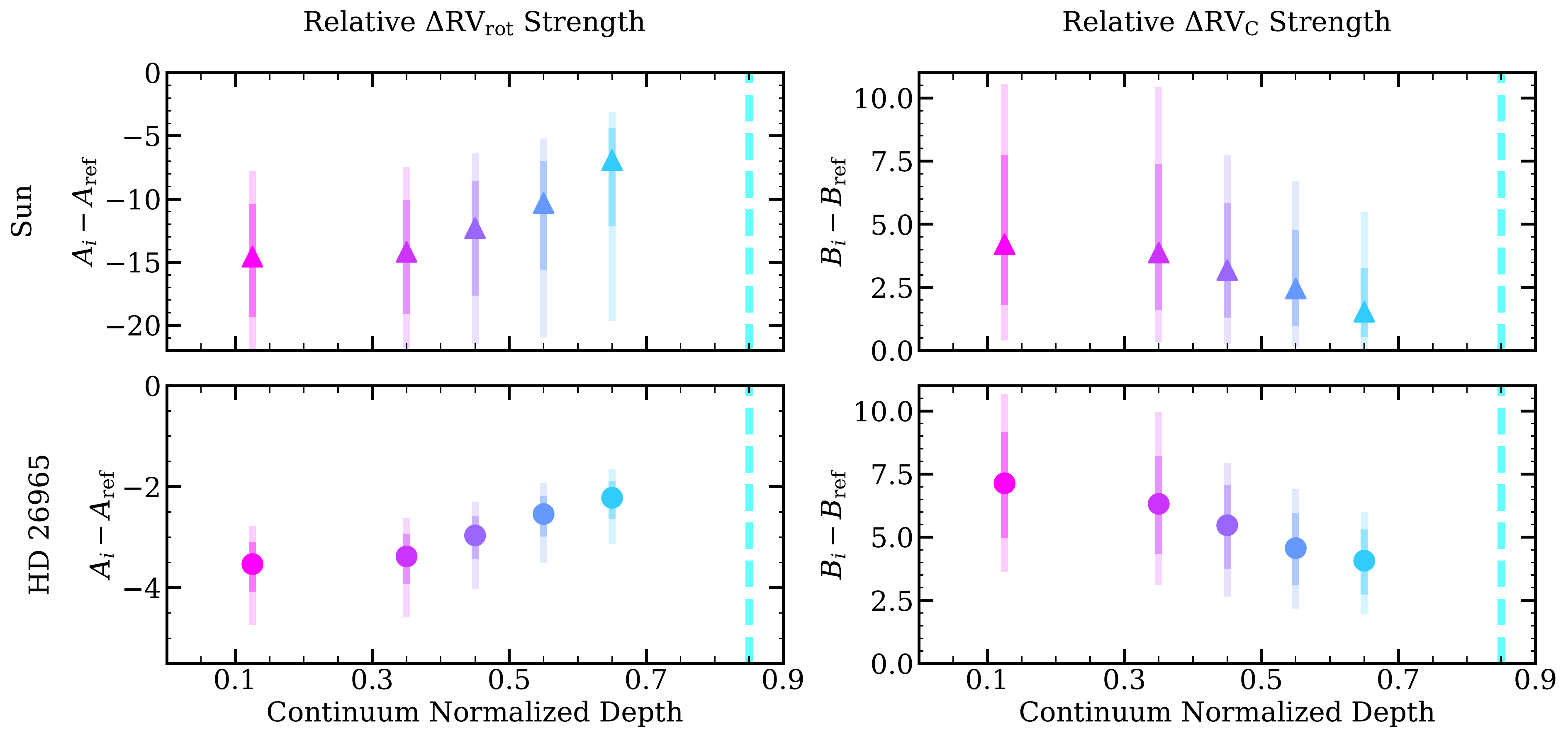}{
\textwidth}{} }
\caption{ 
The RC model maps the relative strengths of the rotation $\Delta \mathrm{RV}_\mathrm{rot}(t)$ and convection $\Delta \mathrm{RV}_\mathrm{C}(t)$ perturbations induced by active regions as a function of continuum normalized depth.
The results for the Sun and {\HDtwo} are presented in the top and bottom rows, respectively.
Left: the strength of the rotation perturbation as a function of depth in the 1--dimensional spectrum (relative to the ``bottom'' spectral slice, see Figure~\ref{fig:RVspot_RVconv_recovery} for a diagram); the rotation perturbation is strongest at heavily absorbed wavelengths, resulting in $A_i-A_\mathrm{ref}<0$.
The central point, dark shaded region, and light shaded region represent the median, $1\sigma$, and $2\sigma$ confidence intervals of the MCMC samples, respectively.
Right: the strength of the convection perturbation $\Delta \mathrm{RV}_\mathrm{C}(t)$ as a function of depth in the spectrum; the convection perturbation is strongest at weakly absorbed wavelengths, resulting in $B_i-B_\mathrm{ref}>0$.
The depth of the reference spectral slice is shown as a dashed-vertical line.
}
\label{fig:versus_depth}
\end{figure*}

\begin{figure*}[t]
\gridline{\fig{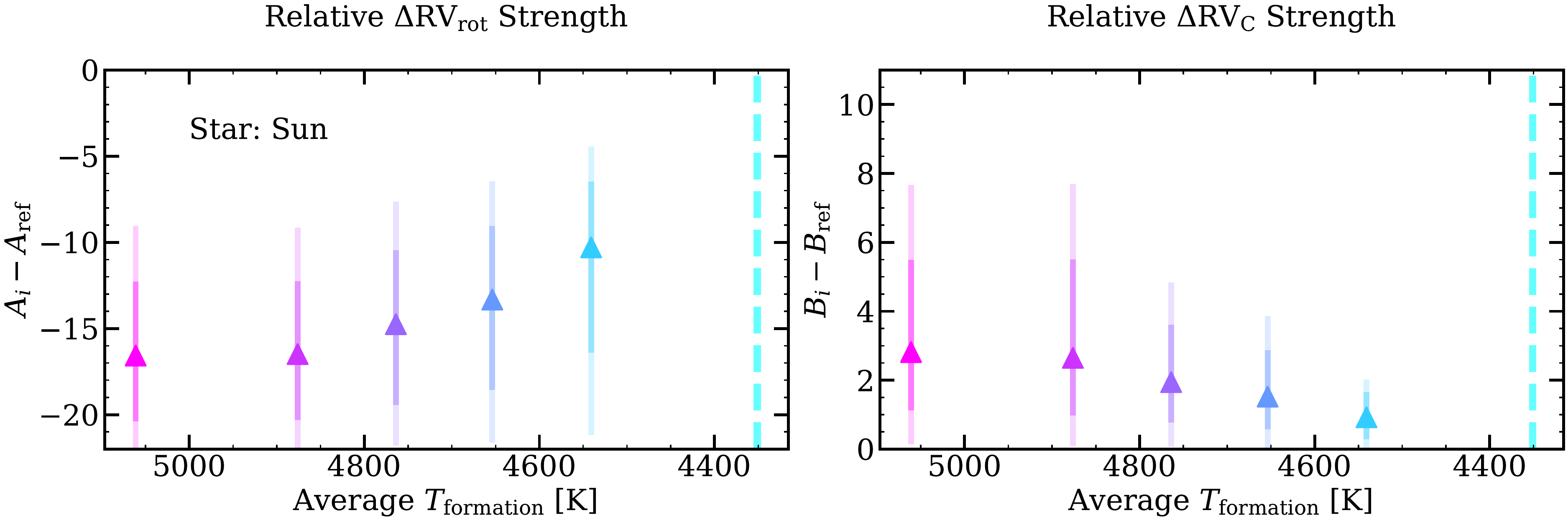}{
\textwidth}{} }
\caption{ 
The contributions of the rotation $\Delta \mathrm{RV}_\mathrm{rot}(t)$ and convection $\Delta \mathrm{RV}_\mathrm{C}(t)$ perturbations as a function of average temperature of formation in the stellar atmosphere.
Using only spectral features well reproduced by the stellar atmosphere model, the NEID solar RVs were jointly fit with the RC model for five choices of $\mathrm{RV}_i(t) - \mathrm{RV}_\mathrm{ref}(t)$.
Left: the relative strength of $\Delta \mathrm{RV}_\mathrm{rot}(t)$ between RV$_i(t)$ and RV$_\mathrm{ref}(t)$ as a function of the average formation temperature; 
the average formation temperature of RV$_\mathrm{ref}(t)$ is shown as a dashed-vertical line.
Right: the relative strength of $\Delta \mathrm{RV}_\mathrm{C}(t)$ between RV$_i(t)$ and RV$_\mathrm{ref}(t)$.
}
\label{fig:versus_temp}
\end{figure*}

We divided the stellar spectrum into six slices (diagrammed in the leftmost panel of Figure~\ref{fig:RVspot_RVconv_recovery}).
For a given observation, we measured the RV of each slice using the pixel by pixel method.
The average RV uncertainties (photon-noise only) for the spectral slices are $9$ and $72$~{\cms}, for the Sun and {\HDtwo}, respectively; the RV uncertainties vary by only a few {\cms} between the slices.
The spectrum averaged line by line RVs were simultaneously fit using each combination of $\mathrm{RV}_i(t)-\mathrm{RV}_\mathrm{ref}(t)$, where RV$_\mathrm{ref}(t)$ corresponds to the ``bottom'' spectral slice (strongest absorption).
Following Section~\ref{sec:detrending}, we assumed a linear relationship between flux and the flux proxy (the depth metric) and let $\delta V_c \kappa /f$ vary as a linear function of time (with slope $m_{\delta V_c \kappa / f}$ and intercept $b_{\delta V_c \kappa / f}$).
The model was optimized via MCMC sampling; uniform priors were adopted on the free--parameters: 
$A_{1,\mathrm{ref}}^{-1},\dots,A_{5,\mathrm{ref}}^{-1}$, $B_{1,\mathrm{ref}},\dots,B_{5,\mathrm{ref}}$, $m_{\delta V_c \kappa / f}$, and $b_{\delta V_c \kappa / f}$.

The NEID solar and {\HDtwo} datasets were each modeled with the joint-fitting procedure.
In addition to detrending the rotation--modulation signal, the joint fits infer the $\Delta \mathrm{RV}_\mathrm{rot}(t)$ and $\Delta \mathrm{RV}_\mathrm{C}(t)$ perturbations.
For a given joint fit, each $\Delta \mathrm{RV}_{i,\mathrm{ref}}(t)$ yields a prediction for $\Delta \mathrm{RV}_\mathrm{rot}(t)$ and $\Delta \mathrm{RV}_\mathrm{C}(t)$.
The perturbations are presented in Figures~\ref{fig:RVspot_RVconv_recovery}~and~\ref{fig:RVspot_RVconv_HD26965}. 
The inferred perturbations show minimal dependence on $\Delta \mathrm{RV}_{i,\mathrm{ref}}(t)$; 
the five detrending models from a given joint fit differ at the $10$~{\cms} level for both the Sun and {\HDtwo}.
After detrending, the average RMS was $1.04$ and $1.15$~{\ms} for  the Sun and {\HDtwo}, respectively.
Relative to the RC fits in Sections~\ref{sec:detrending}~and~\ref{sec:keplerian}, which used two large spectral slices, the increased photon-noise from using many narrower slices only increased the residual RMS by a few {\cms}.

To validate our decomposition of the rotation--modulation signal, we turned to SDO spatially resolved observations of the Sun; described in Section~\ref{sec:obs}.
We calculated the expected ``Sun-as-a-star'' disk-integrated rotation--modulation RV signal from SDO maps of limb-darkening corrected continuum intensity and magnetic field strength, using \texttt{solaster} \citep[][]{Haywood2016, Ervin2022}. 
The solar RC model is compared to the SDO measurements in Figure~\ref{fig:RVspot_RVconv_recovery}.

The rotation and convection perturbations inferred by the RC model are similar to the SDO measurements; the independent measurements from SDO agree with the RC results at the meter--per--second level.
By accurately decomposing the rotation--modulation signal, we anticipate the RC model will enable novel studies of stellar activity. 

\subsection{Stellar activity versus atmospheric height}

\label{sec:vs_atmo}

In addition to recovering the individual components of rotation--modulation---$\Delta \mathrm{RV}_\mathrm{rot}(t)$ and $\Delta \mathrm{RV}_\mathrm{C}(t)$---the RC model also measures the perturbations' strengths as a function of atmospheric height.
This is accomplished via the joint fitting procedure introduced in Section~\ref{sec:phys}.

As diagrammed in Figure~\ref{fig:RVspot_RVconv_recovery}, we divided each spectrum into six slices (i.e., binning by absorption strength). 
For each observation, we measured the RV of each slice via the pixel by pixel method. 
We refer to the bottom slice (maximally absorbed wavelengths) as $\mathrm{RV}_\mathrm{ref}(t)$;
the other five slices are denoted $\mathrm{RV}_i$(t), each referring to successively more absorbed wavelengths (i.e., successively higher in the stellar atmosphere).
The RC model returns the relative strengths of the rotation and convection components between $\mathrm{RV}_\mathrm{ref}(t)$ and all five $\mathrm{RV}_i(t)$.

For the Sun and {\HDtwo}, Figure~\ref{fig:versus_depth} presents the rotation and convection amplitudes as a function of depth in the 1--dimensional spectrum.
The magnitude of the rotation perturbation increases with absorption strength, while the magnitude of the convection perturbation decreases with absorption.
This trend was previously reported for the convective component \citep[e.g.,][]{Gray2005, Cretignier2020, AlMoulla2022} but only marginally detected for the rotation component \citep{AlMoulla2022}. 

Absorption relative to the local continuum is a useful, model--free proxy for atmospheric height, however, this approximation neglects changes in continuum opacity with wavelength.
To more accurately estimate the formation height at a given wavelength, we performed 1--dimensional local thermodynamic equilibrium spectral synthesis with PySME \citep[Spectroscopy Made Easy, ][]{Valenti1996,Piskunov2017,Wehrhahn2023}; our approach followed \cite{AlMoulla2022}, which we briefly describe below.
We only considered the Sun for this analysis. 
The model was initialized with a VALD line list \citep{Piskunov1995,Ryabchikova2015} based on the temperature, metallicity, surface gravity, and line-broadening listed in Table~\ref{tab:obs}.
A range of atmospheric heights contribute to the emergent flux at a given wavelength; spectral synthesis calculates this distribution (Eqn.~\ref{eqn:cumulative_contribution_fn}) as a function of optical depth (or equivalently temperature).
At a given wavelength, we summarized the contribution function with the characteristic temperature $T_\mathrm{formation}$: the temperature such that the normalized cumulative contribution function equals 50\%.
In the photosphere, temperature monotonically decreases with atmospheric height. 
Temperature is therefore a proxy for relative height in the photosphere.
Following \cite{AlMoulla2022}, we do not convert formation temperature to a geometric height; this conversion would require placing our photosphere model in the context of a complete stellar model, potentially introducing additional systematic uncertainties.

With the above procedure, we calculated $T_\mathrm{formation}$ versus wavelength in the synthetic PySME spectrum (for the Sun).
Unfortunately, the synthesized spectrum only partially reproduces the observed stellar spectrum, due to incompleteness in the input VALD line list.
For each spectral line retained by the line by line pipeline, we compared the observed reference spectrum to the synthesized spectrum.
To account for line-depth, the observed and synthetic lines were first normalized from $0$ to $1$.
Lines were then retained if $\sum_{n=1}^N \sqrt{ [I_0(n)-I_S(n) ]^2}/N < 0.2$, where $N$ is the number of pixels in a given line's reference spectrum window, $I_0(n)$ is the reference spectrum normalized flux, and $I_S(n)$ is the normalized synthetic spectrum flux evaluated on the reference spectrum wavelength solution.
Approximately 2200/3700 lines were retained.
For the retained lines, we mapped $T_\mathrm{formation}$ from the synthetic spectrum to the observed reference spectrum via linear-interpolation.

Using the temperature mapped line list, we repeated the RC modeling procedure. 
Figure~\ref{fig:versus_temp} presents the rotation and convection perturbation strengths as a function of average $T_\mathrm{formation}$, for the Sun.
The filtered dataset shows minor differences from our original RC modeling results, however, these changes are to be expected; the temperature mapped line list includes $40\%$ fewer spectral lines than the initial solar dataset and cross-matching with the synthetic spectrum preferentially rejected shallow spectral lines.
We again see the magnitude of the rotation perturbation decreases with temperature, while the strength of the convection perturbation increases with temperature.
Loss of the shallow spectral lines unfortunately reduces the model's constraining power, because shallow spectral lines are more sensitive to the convection perturbation. 
By combining the RC model with spectral synthesis, we are now tracing the perturbation strengths explicitly as a function of formation temperature.

Complicating matters, the rotation perturbation is expected to vary with depth relative to the line core, even if the flux imbalance induced by a spot is constant through the stellar photosphere;
see Section~\ref{sec:RC_model} for details.
This muddles the physical interpretation of Figures~\ref{fig:versus_depth}~and~\ref{fig:versus_temp}: 
is the strength of the rotation perturbation dependent on continuum normalized depth because of an underlying dependence on formation height, an artifact of the normalization approach (continuum versus line core), or a combination of the two?
This ambiguity could be addressed by simultaneously modeling the dependence of the rotation perturbation on depth in the line profile and depth relative to the local continuum. 
However, modeling the effects of active regions as a function of depth in the line profile would require fitting for the sizes and positions of individual activity complexes, which falls beyond the scope of this study \citep[e.g.,][]{DiMaio2023}.

Lastly, we consider whether the inferred trends between perturbation strength and absorption strength are statistically significant.
To address this, we repeated the joint fitting process but assumed the same perturbation strengths for each choice of $\mathrm{RV}_i(t) - \mathrm{RV}_\mathrm{ref}(t)$.
For simplicity, we did not enforce the temperature mapped line list (i.e., we used the same spectrum averaged RVs and depth binned RV$_i(t)$ used in Section~\ref{sec:phys}).
When $B_{i,\mathrm{ref}}$ does not vary with $i$, it remains degenerate with $\delta V_c \kappa /f$. 
We therefore returned to the simplified detrending model introduced in Section~\ref{sec:detrending}, which bypasses this degeneracy at the cost of no longer separating the rotation and convection perturbations.
For the Sun and {\HDtwo}, the two models---i.e., allowing the perturbation strengths to vary with depth versus held constant with depth---yield residual RMS within $10$~{\cms} of each other.
In terms of relative maximum log-likelihood---i.e., the difference in maximum likelihood achieved by two competing models $\Delta \mathcal \ln {\hat{L}} = \ln {\hat{L}_\mathrm{Model~1}}-\ln {\hat{L}_\mathrm{Model~2}}$---the more flexible model is favored at $\Delta \mathcal \ln {\hat{L}} \sim 1.5$ and $6$, for the Sun and {\HDtwo}, respectively. 
Since the more complex model has 12 free--parameters ($A_{1,\mathrm{ref}},\dots,A_{5,\mathrm{ref}}$, $B_{1,\mathrm{ref}},\dots,B_{5,\mathrm{ref}}$, $m_{\delta V_c \kappa / f}$, and $b_{\delta V_c \kappa / f}$), while the simpler model has 3 ($m_\alpha$, $b_\alpha$, and $\beta$), the simpler model is preferred by the Bayesian information criterion: $\mathrm{BIC} = k \ln (n) - 2 \ln \hat{\mathcal{L}}$, where $k$ is the number of free--parameters, $n$ is the number of data-points, and $\hat{\mathcal{L}}$ is the maximal likelihood. 

Statistical evidence of the perturbation amplitudes varying with depth in the 1--dimensional spectrum is lacking.
However, two factors suggest the trend may exist but is presently obscured by systematics. 
Firstly, previous observational studies found a relationship between a line's average formation height in the stellar atmosphere and (i) absolute convective blueshift \citep[commonly known as the third-signature of stellar granulation,][]{Gray2009,Liebing2021} and (ii) the convection---and potentially the rotation---perturbation(s) induced by an active region \citep{Cretignier2020,AlMoulla2023}.
Secondly, the rotation and convection perturbations inferred by the RC model agree with independent measurements from SDO.
This suggests the RC model is properly tracing and decomposing the rotation--modulation signal. 
Future work addressing the systematics noise floor is warranted.

The RC model---a reframing of the spot model underlying the FF$^\prime$ method---successfully detrends the rotation--modulation RV signal and opens new avenues for characterizing stellar activity.
Using only the 1--dimensional spectrum, the revised spot model accurately recovered the underlying rotation and convection perturbations induced by spots and their associated magnetic regions; see Section~\ref{sec:phys}.
Here, we explored an additional facet of the RC model: tracing the strengths of the rotation and convection perturbations as a function of depth in the 1--dimensional spectrum.
For both the Sun and standard star {\HDtwo}, the magnitude of the rotation perturbation \textit{increased} with absorption strength, while the magnitude of the convection perturbation \textit{decreased} with depth in the spectrum.

\section{Discussion}
\label{sec:discussion}

In this study, we developed and tested the RC model: a method of detrending and characterizing the anomalous rotation--modulation RV signal.
The RC model decomposes rotation--modulation into two parts---rotation and convection---and considers how these components change with height in the stellar atmosphere. 
Here we diagnose the RC model's performance and outline avenues for future study.  

For the Sun and the well-studied standard-star {\HDtwo}, the RC model successfully detrends the rotation--modulation RV signal, lowering the measured RV variability from $\gtrsim 2$ to $1$~{\ms}. 
The residual RV variation was found to be independent of observing cadence; for {\HDtwo}, we considered HARPS, EXPRES, and NEID datasets, which differ by an order of magnitude in the frequency of observations but resulted in comparable RV RMS values.

The RC model only considers RV variability from rotation--modulation. 
Other sources of anomalous RV signals (e.g., granulation) are unaccounted for in this framework.
For context, we consulted the Gaussian process covariance kernels of \cite{Luhn2023}, which predict the RV variability due to p-mode oscillations and granulation as a function of stellar type and survey design; the effects of supergranulation were not included.
For the solar dataset, the predicted residual noise level is $\sqrt{ 3.7^2 + 11.5^2 + 30^2 + 2^2 } \approx 30$~{\cms}, for p--modes, granulation, instrumental systematics, and photon-noise, respectively. 
For the NEID {\HDtwo} dataset, the predicted residual noise level is $\sqrt{ 24^2 + 21^2 + 30^2 + 16^2 } \approx 45$~{\cms}. 
The expected jitter from unaccounted for sources of stellar variability falls short of the observed residuals, leaving room for future improvement, which we explore below; undetected Keplerian signals or underestimation of the p-mode oscillations and granulation jitter are also potential culprits. 

The RC method assumes the same spot model underlying the FF$^\prime$ method. 
The FF$^\prime$ method expresses the RV signature of an active region in terms of the host star's flux variations, the spots' sizes and positions on the stellar surface, and the strength of the associated magnetic regions \citep{Aigrain2012}.
Evaluating the FF$^\prime$ model relies upon numerically differentiating the flux (or flux proxy) timeseries.
In contrast, the inputs to the RC model are flux (or a flux proxy) and the difference between the RVs drawn from higher in the stellar atmosphere (strongly absorbed wavelengths) and those drawn from deeper in the atmosphere (weakly absorbed wavelengths);
the RC model does not require numerical differentiation.
Since the RC and FF$^\prime$ methods rely on the same underlying model, differences between their detrending results reveal systematics from either numerical differentiation or our relative formation height approach.
On datasets with high enough temporal sampling to numerically differentiate the flux timeseries, the RC and FF$^\prime$ methods yield near identical results; see Figures~\ref{fig:solar_summary}~and~\ref{fig:rv_comparison}.
The detrending results are therefore independent of the method of applying the simplified spot model (i.e., numerical differentiation versus relative formation height), with the caveat that the RC method has far more forgiving cadence requirements.

In order to evaluate the RC model, we made the simplifying approximations that (i) the average properties of the host star's surface features are constant or linearly varying with time and (ii) an activity indicator is an adequate proxy of the host star's flux.
These choices only apply to how the RC model was fit, not the underlying formalism.
Future work relaxing these approximations is warranted. 
For instance, the spot fraction could significantly vary over a few stellar rotation periods or there may be a time lag between an activity indicator and stellar flux \citep{Burrows2024}.
To account for changes in the spot properties as a function of time, the free--parameters of the RC model could be treated as Gaussian process kernels.
Quantifying the relation between stellar flux and different activity indicators---e.g., H$\alpha$, $\log R^\prime_\mathrm{HK}$, FWHM of the CCF---could help reign in systematic uncertainties associated with the use of a flux proxy.
Instead of adopting a single flux proxy, the host star's flux timeseries could also be approximated as a Gaussian process conditioned on several activity indicators.

Detrending performance may also be improved by accounting for additional sources of stellar variability.
In this work, we revisited the simplified spot model from the perspective of formation height in the stellar atmosphere.
This approach is well poised for jointly modeling the effects of granulation and rotation--modulation.
Since rising granules contribute more observable light than the falling granules, disk-integrated spectral lines can display ``C''-shaped bisectors \citep{Stathopoulou1993,Gray2008}.
Prior studies of the line distortions induced by granulation often invoke magnetohydrodynamic simulations of the photosphere \citep[e.g.,][]{Cegla2019}.
Recently, \cite{Palumbo2022} introduced the GRanulation And Spectrum Simulator (\texttt{GRASS})---a tool for generating timeseries of granulation perturbed synthetic spectra---based on the solar observations of \cite{LB2019}.
Jointly modeling the line distortions of granulation and rotation--modulation would be a significant advancement.

The RC model has proven a valuable tool for both detrending and characterizing rotation--modulation.
For both the Sun and {\HDtwo}, the RC model closely matches  FF$^\prime$ detrending, without the need for high cadence observations. 
Moving forward, there are several areas ripe for development, including (i) accounting for the growth and decay of the active regions, (ii) exploring improved flux proxies, and (iii) simultaneously modeling rotation--modulation and granulation. 

\section{Conclusions}
\label{sec:conclusions}

Star spots and their associated magnetic regions induce RV perturbations through two effects: (i) the spot induces a flux imbalance between the rotationally red- and blueshifted hemispheres and (ii) the magnetic region locally suppresses convective blueshift.
As the star rotates and the active regions cross the stellar surface, these perturbations generate a time-dependent RV signal.
Rotation--modulation remains a significant challenge in the detection and characterization of extrasolar planets; 
rotation--modulation has previously been mistaken for Keplerian motion \citep{Lubin2021}, and with periods on the order of days, these signals cannot easily be averaged out with a targeted observing strategy.

In this paper, we revisited the physical picture of rotation--modulation. 
Previously, \cite{Aigrain2012} parameterized the expected RV signal from rotation--modulation in terms of the host star's flux variations and the active regions' physical properties. 
This approach---commonly known as FF$^\prime$---relies upon high temporal sampling to approximate the first time derivative of the host star's flux (or a flux proxy).
Our motivation was to reframe the spot model such that the physics of stellar activity are readily explorable, while maintaining accessibility (i.e., defining the model in terms of common RV data-products) and applicability (i.e., relaxing the requirement on high cadence observations).

We developed the Rotation--Convection (RC) method. 
Previous observational studies pointed towards the rotation and convection perturbations induced by an active region varying with atmospheric height \citep{Gray2009, Liebing2021, Cretignier2020,AlMoulla2023}.
The RC method expands upon the simplified spot model, by considering how the rotation and convection perturbations induced by an active region change with height in the stellar atmosphere.
Unlike the FF$^\prime$ approach, the RC model does not rely upon the first time derivative of flux and therefore has limited cadence requirements. 
Evaluating the RC model only requires the 1--dimensional spectrum or the cross correlation function.

We applied the RC model to NEID solar data, as well as HARPS, EXPRES, and NEID observations of the standard-star {\HDtwo}. 
In each case, the RC model successfully detrends the rotation--modulation signal, lowering the RV variability from $\gtrsim 2$ to $1$~{\ms}.
The RC model's detrending performance was independent of observing cadence; the observing cadence differs by an order of magnitude between the three {\HDtwo} datasets.
Injection tests confirmed the RC model preserves Keplerian information.
For the NEID solar and {\HDtwo} datasets, where the temporal sampling is sufficient to apply FF$^\prime$ detrending (i.e., numerical differentiation of a flux proxy timeseries),
the RC and FF$^\prime$ models yield near identical results, even though the RC model has far more forgiving cadence requirements.

By design, the RC model both detrends and \textit{characterizes} the rotation--modulation signal. 
We confirmed the model accurately recovers and separates the rotation and convection RV components;
this decomposition was validated by independently measuring the components of rotation--modulation with high-spatial-resolution Solar Dynamics Observatory observations \citep{Pesnell2012, Scherrer2012, Haywood2016, Ervin2022}.
The RC model also traces the amplitude of the rotation and convection perturbations as a function of depth in the 1--dimensional spectrum (i.e., absorption strength).
For both the Sun and standard star {\HDtwo}, we found the amplitude of the rotation perturbation \textit{increased} with absorption strength, while the amplitude of the convection perturbation \textit{decreased} with absorption strength.
These trends can be interpreted in terms of height in the stellar atmosphere, given the correlation between depth relative to the local continuum and atmospheric height;
however, the rotation perturbation is expected to vary with depth relative to the line core, even if the active regions' flux contrast is constant through the photosphere, introducing a level of ambiguity. 

Using only standard RV data-products and with limited cadence requirements, the RC model successfully detrends and characterizes the rotation--modulation RV signal. 
We anticipate the RC model will inform future methods of robust RV detrending and open the door to future studies of stellar atmospheres.

\acknowledgments
Some data presented were obtained by the NEID spectrograph built by Penn State University and operated at the WIYN Observatory by NOIRLab, under the NN-EXPLORE partnership of the National Aeronautics and Space Administration and the National Science Foundation. Based in part on observations at the Kitt Peak National Observatory, managed by the Association of Universities for Research in Astronomy (AURA) under a cooperative agreement with the National Science Foundation. WIYN is a joint facility of the University of Wisconsin–Madison, Indiana University, NSF’s NOIRLab, the Pennsylvania State University, Purdue University, University of California, Irvine, and the University of Missouri. The authors are honored to be permitted to conduct astronomical research on Iolkam Du’ag (Kitt Peak), a mountain with particular significance to the Tohono O’odham. 

Data presented herein were obtained at the WIYN Observatory from telescope time allocated to NN-EXPLORE through the scientific partnership of the National Aeronautics and Space Administration, the National Science Foundation, and the National Optical Astronomy Observatory. The research was carried out, in part, at the Jet Propulsion Laboratory, California Institute of Technology, under a contract with the National Aeronautics and Space Administration (80NM0018D0004). 

These results made use of the Lowell Discovery Telescope at Lowell Observatory. Lowell is a private, non-profit institution dedicated to astrophysical research and public appreciation of astronomy and operates the LDT in partnership with Boston University, the University of Maryland, the University of Toledo, Northern Arizona University and Yale University.

We gratefully acknowledge the contributions of the EXPRES team.  The EXPRES team acknowledges support for the design and construction of EXPRES from NSF MRI-1429365, NSF ATI-1509436 and Yale University. 

This work has been carried out within the framework of the NCCR PlanetS supported by the Swiss National Science Foundation under grants 51NF40\_182901 and 51NF40\_205606. This project has received funding from the European Research Council (ERC) under the European Union’s Horizon 2020 research and innovation program (grant agreement SCORE No. 851555).

JS acknowledges support by the National Science Foundation Graduate Research Fellowship Program under Grant DGE-2039656. 

GS acknowledges support provided by NASA through the NASA Hubble Fellowship grant HST-HF2-51519.001-A awarded by the Space Telescope Science Institute, which is operated by the Association of Universities for Research in Astronomy, Inc., for NASA, under contract NAS5-26555.

EBF acknowledges support by Heising-Simons Foundation Grant \#2019-1177 and NASA Grant \# 80NSSC21K1035.
The Center for Exoplanets and Habitable Worlds is supported by Penn State and its Eberly College of Science.

\bibliography{paper}%

\end{document}